\documentstyle[preprint,eqsecnum,aps]{revtex}
\begin{document}
\draft
\title
{
The mechanism of spin and charge separation in one dimensional quantum
antiferromagnets
}
\author{Christopher Mudry and Eduardo Fradkin}
\address{Physics Department, University of Illinois at Urbana-Champaign,
1110 West Green Street, Urbana, Illinois 61801-3080}
\date{\today}
\maketitle

\begin{abstract}
We reconsider the problem of separation of spin and charge in one dimensional
quantum antiferromagnets.
We show that spin and charge separation in one dimensional strongly
correlated systems cannot be described by the slave boson or fermion
representation within any perturbative treatment of the interactions
between the slave holons and slave spinons. The constraint of single occupancy
must be implemented exactly.
As a result the slave fermions and bosons are not part of the
physical
spectrum.
Instead, the excitations which carry the separate spin and charge
quantum numbers are solitons.
To prove this {\it no-go} result,
it is sufficient to study the pure spinon sector in the slave boson
representation. We start with a short-range RVB spin liquid mean-field
theory for the frustrated antiferromagnetic spin-${1\over2}$ chain.
We derive an effective theory for the fluctuations of the Affleck-Marston
and Anderson order parameters.
We show how to recover the phase diagram as a function of the
frustration by treating the fluctuations non-perturbatively.

\end{abstract}

\narrowtext

\section{Introduction}
\label{sec:Introduction}

Since the discovery of high $T^{\ }_c$ superconductivity,
there has been a lot of interest in the $t-J$ model
\cite{tjmodel}
for low dimensional lattices. It is known that for a linear chain,
the $t-J$ model displays the phenomenon of {\it spin and charge separation}
for strong enough electronic correlations
\cite{tJisluttingernum,hubbisluttingeranal,tJisluttingeranal}
and belongs to the class of Luttinger liquids \cite{Haldane 1980}.
The spin and charge excitations which carry separately the
elementary charge of the electron are  {\it solitons}
in a Luttinger liquid. This is most easily seen using a
Jordan-Wigner representation of the spin-${1\over2}$
in the context of the Heisenberg chain.
Much effort has been devoted to prove the existence of this
phenomenon in higher dimensional $t-J$ models (most significantly in two
dimensions in view of its possible relevance to high $T_c$ superconductivity).

One approach to this issue has been to think of the {\it band electron}
as a {\it point-like} or {\it local}
bound state of two constituents: one, the {\it slave holon},
carrying the electronic charge, the other, the {\it slave spinon},
carrying the electronic
spin quantum number \cite{slavefermibose}.
The slave holon and spinon
constituents are held (glued) together by a {\it strongly} fluctuating gauge
field. The hypothesis behind this picture is that,
by some {\it deconfining} mechanism, it is favorable for the system to
liberate the electron constituents, i.e., the energy cost for breaking
the local bound state is finite (possibly zero).
If this is so, it is then natural as a first step to approximate
the interacting system of holons and spinons by free holons and free
spinons with self-consistently determined renormalized kinetic energy
scales and then to include perturbatively the gauge fluctuations.
This is the content of all the
mean-field theories for slave holons and spinons
which freeze the strong gauge fluctuations between the holons and spinons
in order to describe the separation of spin and charge.
For lack of an alternative, this strategy has been widely used to implement
spin and charge separation in two dimensions.
On the other hand, this picture is not useful
if the holons and spinons are always constrained
to form bound states on the shortest possible scale, i.e.,
the lattice spacing.

The holon and spinon carry complementary statistics.
In the slave boson scheme,
the holon is a boson while the spinon is a fermion
\cite{Baskaran 1987,Zou 1988}.
In the slave fermion scheme, the holon is a fermion and the
spinon is a boson \cite{Arovas 1988,Read 1989,Flensberg 1989}.
For the $t-J$ model, the mean-field predictions for both schemes
differ qualitatively. This is not surprising since the slave boson
scheme is most appropriate  when the holes are moving in the background
of a {\it spin liquid} whereas the slave fermion scheme is convenient in
the presence of strong antiferromagnetic correlations.
In one dimension and at the mean-field level
neither the slave boson nor the slave fermion scheme describes
a Luttinger liquid \cite{Feng 1993}.
Moreover, this failure of the mean-field theory in one dimension
persists even if the self-energy corrections
to the holons and spinons
propagators due to the gauge fields
are included (as shown by Feng {\it et al} \cite{Feng 1993}).
One might wonder then if vertex corrections due to the gauge fields
are sufficient to restore
the characteristic features of a Luttinger liquid and if not,
how one recovers the Luttinger liquid.

In this paper, we show that it is {\it not possible}
to obtain Luttinger liquid behavior
by including {\it perturbatively}
gauge fluctuations around the mean-field Ansatz for the holon-spinon system,
the reason being that deconfinement is never allowed in one dimension.
The separation of spin and charge in the one dimensional
$t-J$ model is due to a {\it topological mechanism} which is
{\it qualitatively different}
from the mechanism of deconfinement.
In one dimension, the charge elementary excitations and the spin elementary
excitations are {\it solitons}.
This {\it no-go} result is important
in view of the fact that we have shown that
a spin liquid proposed by Wen \cite{Wen 1991}
{\it can} support the deconfinement of the slave spinons
in two-dimensional (and possibly higher) strongly correlated systems
\cite{Mudry 1994}.

We are thus confronted with the following situation.
On the one hand, spin and charge separation occurs in
one dimensional strongly correlated electronic systems
but it {\it cannot} be described by the deconfinement of local bound states
of holons and spinons constituting the band electrons.
On the other hand, the same local bound states of holons and spinons
can break up in higher dimensional strongly correlated systems,
thus achieving a {\it different} realization of spin and charge separation.
It is not known at the present if an analogy of
the topological mechanism of spin and charge
separation is available in higher dimensional systems. It appears that
dimensionality is crucial and it could well be that the
one dimensional mechanism is not available in higher dimensions.

The essence of our {\it no-go} result is that the lower
(space-time) critical dimension for the deconfinement of
a pure gauge theory with a {\it discrete} symmetry group is $2+1$
\cite{Wegner 1971,Mudry 1994}. Hence, it applies to either the slave
boson or slave fermion scheme. Moreover, it is sufficient to consider
the pure spinon sector and to prove that the spinons can never deconfine.
This we do by studying the Heisenberg chain for
spin-${1\over2}$ with
antiferromagnetic nearest and next-nearest neighbor couplings.
We choose to represent the spin-${1\over2}$ by fermionic spinons.
In other words, we use the slave boson representation
of the $t-J_1-J_2$ model at half-filling.

Our motivation for this choice is three-fold.
First, in two dimensions, deconfinement of the slave spinons
is made possible by the opening of an energy gap in the spinon spectrum
as a result of {\it frustration}, i.e., if the spinon ground state
describes a {\it short-range} spin liquid.
Second, many properties of the frustrated Heisenberg chain for
spin-${1\over2}$ are well understood.
The exact ground state and low
energy excitation spectrum are known in the limits
${J_2\over J_1}=0$ \cite{Faddeev 1981},
${J_2\over J_1}={1\over2}$ \cite{j2/j1=0.5},
and
${J_2\over J_1}=\infty$.
Haldane has constructed the zero-temperature phase diagram
as a function of not too large frustration ${J_2\over J_1}$
\cite{Haldane 1982}.
The parent $t-J_1-J_2$ model has been studied
numerically \cite{Ogata 1991}.
Finally, the model close to the limit ${J_1\over J_2}=0$
resembles the double spin-${1\over2}$ chain problem which has recently
received much attention \cite{doublechain} in connection with the single
rung $t-J$ ladder \cite{t-Jladder}.
Thus, {\it the frustrated Heisenberg chain for spin-${1\over2}$ is an ideal
model to understand the mechanism by which
a separation of spin and charge unrelated to deconfinement
is realized in the gauge field approach}.

To construct a spin liquid, we choose in Sec. \ref{sec:equiv}
to rewrite the spin problem as a SU(2) lattice gauge theory coupling
fermionic spinons with gauge fields. Our choice for a spin liquid
ground state is given in Sec. \ref{sec:s-RVBMF}. The mean-field theory
for the spinons is described and an effective field theory which
includes {\it all} the smooth fluctuations of the order parameters
characterizing the spin liquid is derived.
We treat the quantum fluctuations of the order parameters {\it exactly}
in Sec.  \ref{sec:Nonabelianboso}. We show explicitly how all the
predictions of the mean-field theory are modified  by the quantum
fluctuations. All the mean-field excitations are {\it removed},
quantum criticality for small frustration ${J_2\over J_1}$ is restored.
At criticality, the effective quantum field theory is shown to be the
level $k=1$ Wess-Zumino-Witten theory in agreement with Affleck and Haldane
\cite{Affleck 1987,Affleck 1986}.
The gapless excitations in the critical regime are identified
with the Jordan-Wigner topological excitations.
We show in Sec. \ref{sec:Dimerization} how
the second order phase transition from a gapless spin liquid
to a dimerized phase is induced by frustration in our quantum field theory.
Finally, we briefly discuss an effective quantum field theory for
two weakly interacting antiferromagnetic Heisenberg chains.

\section{The equivalence of the Heisenberg model for spin-${1\over2}$
to a SU(2) lattice gauge theory}
\label{sec:equiv}

We start with the Heisenberg model for spin-${1\over2}$
\begin{equation}
H=\sum_{\langle ij\rangle}\; J_{ij}\; \vec S_i\cdot\vec S_j,
\label{Heisenberghamiltonian}
\end{equation}
where $\langle ij\rangle$
is an ordered pair of sites on an arbitrary lattice $\Lambda$,
$J_{ij}$ are real coupling constants. The slave fermion (or {\it spinon})
representation of the spin-${1\over2}$ operators is
\begin{equation}
\vec S_i=
{1\over2}\;
s^{\dag}_{i\alpha}\;
\vec\sigma^{\   }_{\alpha\beta}\;
s^{\   }_{i\beta },
\label{spininspinonbasis}
\end{equation}
$\vec\sigma$ being the Pauli matrices and the $s^{\ }_i$'s obeying fermionic
anticommutation relations.
The spinon representation, Eq.\ (\ref{spininspinonbasis}),
for the spin-${1\over2}$
degrees of freedom must be supplemented with {\it any} of the three constraints
\begin{equation}
\openone=
s^{\dag}_{i\alpha}\;
\delta^{\   }_{\alpha\beta}\;
s^{\   }_{i\beta }\;
\Leftrightarrow\;
0\;=\; s^{\dag}_{i\uparrow}\; s^{\dag}_{i\downarrow}\;
\Leftrightarrow\;
0\;=\; s^{\   }_{i\downarrow}\; s^{\   }_{i\uparrow}
\label{singleoccupancy}
\end{equation}
for all sites $i$ of the lattice.

{}From the fully symmetric tensor $\delta_{\alpha\beta}$
and the fully antisymmetric tensor
$\epsilon_{\alpha\beta}$
of SU(2), the two bilinear forms
\begin{equation}
\chi^{\dag}_{ij}\; =\;
s^{\dag}_{i\alpha}\;
\delta^{\   }_{\alpha\beta}\;
s^{\   }_{j\beta }
\end{equation}
and
\begin{equation}
\eta^{\dag}_{ij}\; =\;
s^{\dag}_{i\alpha}\;
\epsilon^{\   }_{\alpha\beta}\;
s^{\dag}_{j\beta },
\end{equation}
can be used to describe a singlet pairing of the two
spin-${1\over2}$
located on site $i$ and $j$, respectively
\cite{Affleck 1988,Baskaran 1987}.
Indeed, the identity
\begin{equation}
\vec S_i\cdot\vec S_j\; =\;
-{1\over4}\; \eta^{\dag}_{ij}\; \eta^{\   }_{ij}\; -\;
 {1\over4}\; \chi^{\dag}_{ij}\; \chi^{\   }_{ij}\; +\;
 {1\over4}\; \openone
\end{equation}
holds in the Hilbert space of one spinon per site.
A {\it spin liquid} which, by definition,
should not show any long range magnetic order, implies,
in the spinon picture, the exponential decay with separation $|i-j|$
of the ground state
expectation values
$\langle\eta^{\dag}_{ij}\rangle$ or
$\langle\chi^{\dag}_{ij}\rangle$.

The dynamics of these bilinear forms can be obtained from the
partition function
\begin{eqnarray}
Z\;=&&
\;
\int{\cal D}\;\left[\vec a_{\hat{\rm o}}\right]\;
\int{\cal D}\;\left[ s^{*}\right]
{\cal D}\;\left[s^{\ }\right]\;
e^{+{\rm i}\int dt\; L'}
\end{eqnarray}
where the lattice Lagrangian is
\begin{eqnarray}
L'\;=&&
\;\sum_i\; s^{*}_{it\alpha}\;{\rm i}\partial^{\ }_t\; s^{\ }_{it\alpha}
\nonumber\\
-&&\;\sum_i\;
\left(
{1\over2}\; a^-_{\hat{\rm o}it}\; \eta^{* }_{iit}+
{1\over2}\; a^+_{\hat{\rm o}it}\; \eta^{\ }_{iit}+
{1\over2}\; a^3_{\hat{\rm o}it}\; (\chi^{*}_{iit}-1)
\right)
\nonumber\\
+&&\;\sum_{\langle ij\rangle}\;
{J_{ij}\over4}\;
\left(
\eta^{*}_{ijt}\eta^{\ }_{ijt}
\;+\;
\chi^{*}_{ijt}\chi^{\ }_{ijt}
\right).
\label{firstlatticelagrangian}
\end{eqnarray}
The integration over the Lagrange multipliers
\begin{equation}
\pmatrix
{
a^-_{\hat{\rm o}it}\cr
a^+_{\hat{\rm o}it}\cr
a^3_{\hat{\rm o}it}\cr
}\;
=\;
\pmatrix
{
{1\over2}\;(a^1_{\hat{\rm o}it}\;-\;{\rm i}\ a^2_{\hat{\rm o}it})\cr
{1\over2}\;(a^1_{\hat{\rm o}it}\;+\;{\rm i}\ a^2_{\hat{\rm o}it})\cr
a^3_{\hat{\rm o}it}\cr
}
\end{equation}
enforces the constraint of single occupancy on the spinon Hilbert space
in a redundant way. But this redundancy allows for the mapping of
Eq.\ (\ref{firstlatticelagrangian})
into the Lagrangian of a SU(2) lattice gauge theory
\cite{Hsu 1988}, if
a Hubbard-Stratanovich transformation with respect to the composite fields
$\eta$ and $\chi$ is performed first.

With a particle-hole transformation of the spinons,
our final Lagrangian will then take the form \cite{Dagotto 1988}
\cite{Baskaran 1989}
\begin{eqnarray}
L\; =&&\;
\sum_i\;
\psi^{* }_{it}\;
\left(\;
{\rm i}\partial^{\ }_t-A_{\hat{\rm o}it}\;
\right)\;
\psi^{\ }_{it}
\label{finallatticelagrangian}\\
-&&\;
\sum_{\langle ij\rangle}{J_{ij}\over4}
\left[
|{\rm det}\; W^{\   }_{ijt}|+
(
\psi^{* }_{it}W^{\   }_{ijt}\psi^{\ }_{jt}+
\psi^{* }_{jt}W^{\dag}_{ijt}\psi^{\ }_{it}
)
\right].\nonumber
\end{eqnarray}
Here, the $\psi$'s are \cite{particle-hole}
\begin{equation}
\psi_{it}=
\pmatrix
{
s^{\   }_{it\uparrow  }\cr
s^{*   }_{it\downarrow}\cr
}.
\label{particleholetrs}
\end{equation}
The $A_{\hat{\rm o}}$'s
belong to the fundamental representation of the su(2) Lie algebra
\begin{equation}
A_{\hat{\rm o}it}\;=\; {1\over2}\; \vec a_{\hat{\rm o}it}\; \cdot\; \vec\sigma.
\end{equation}
Finally, the $W$'s are 2$\times$2 matrices of the form
\begin{equation}
W^{\   }_{ijt}=
\pmatrix
{
-X^{\ }_{ijt}&-E^{\ }_{ijt}\cr
-E^{* }_{ijt}&+X^{* }_{ijt}\cr
},
\label{spatiallinks}
\end{equation}
which satisfy
\begin{equation}
W^{\   }_{ijt}\;=\;W^{\dag}_{jit}.
\end{equation}
The entries $E$ and $X$ of the $W$'s are the Hubbard-Stratanovich degrees
of freedom associated with the spinon bilinears $\eta$ and $\chi$,
respectively.

The lattice Lagrangian in
Eq.\ (\ref{finallatticelagrangian})
is left unchanged by the local
gauge transformations
\begin{eqnarray}
&&
\psi^{\   }_{it}\;\rightarrow\;
\psi'      _{it}\;=\; U^{\   }_{it}\; \psi^{\   }_{it},
\nonumber\\
&&
A_{\hat{\rm o}it}\;\rightarrow\;
A_{\hat{\rm o}it}'\;=\;
U^{\   }_{it}\;
A_{\hat{\rm o}it}\;
U^{\dag}_{it}\;+\;
\left(
{\rm i}\partial^{\ }_t\;
U^{\   }_{it}
\right)
U^{\dag}_{it},
\nonumber\\
&&
W^{\   }_{ijt}\;\rightarrow\;
W'      _{ijt}\;=\;
U^{\   }_{it}\;
W^{\   }_{ijt}\;
U^{\dag}_{jt},
\label{latticegaugeinv}
\end{eqnarray}
for all $U^{\   }_{it}\ \in$ SU(2).
This local symmetry will be called a {\it color symmetry}. It is a different
symmetry from the one generated by global spin rotations. Indeed, under
the particle-hole transformation Eq.\ (\ref{particleholetrs}),
the spin-${1\over2}$ operators of Eq.\ (\ref{spininspinonbasis})
are mapped into
\begin{eqnarray}
&&
S^1_i\;=\;
+{1\over2}\;
\left(\;
\psi^{\dag}_{i1}\;\psi^{\dag}_{i2}\;+\;
\psi^{\   }_{i2}\;\psi^{\   }_{i1}\;
\right)\;\equiv \;
+{1\over2}\;
\left(\;
b^{\dag}_i+b^{\   }_i
\right),
\nonumber\\
&&
S^2_i\;=\;
-{{\rm i}\over2}\;
\left(\;
\psi^{\dag}_{i1}\;\psi^{\dag}_{i2}\;-\;
\psi^{\   }_{i2}\;\psi^{\   }_{i1}\;
\right)\;\equiv \;
-{{\rm i}\over2}\;
\left(\;
b^{\dag}_i-b^{\   }_i
\right),
\nonumber\\
&&
S^3_i\;=\;
+{1\over2}\;
\left(\;
\psi^{\dag}_{i}\;\psi^{\   }_{i}\;-\;1
\right)\;\equiv \;
+{1\over2}\;
\left(\;
m^{\   }_i-1
\right).
\label{spininpsibasis}
\end{eqnarray}

The bilinears $b$ and $m$ defined above are left unchanged by the local
color transformation Eq.\ (\ref{latticegaugeinv}) and thus the Heisenberg
interaction explicitly transforms like a color singlet
when expressed in terms of the $\psi$'s.
Spin-spin correlations can be obtained from the generating functional
\begin{eqnarray}
Z[\ \vec J\ ]\;=&&\;
\int{\cal D}\ \left[W^{\dag}\right]{\cal D}\ \left[W\right]
\int{\cal D}\ \left[\vec a_{\hat{\rm o}}\right]
\int{\cal D}\ \left[\psi^{* }\right]{\cal D}\left[\psi^{\ }\right]\;
e^
{
+{\rm i}\int dt
\left(\
L\ +\ \sum_i\ \vec J_i\cdot \vec S_i\
\right)
}
\end{eqnarray}
where the source term is written in terms of
the bilinears $b$ and $m$ defined by
Eq.\ (\ref{spininpsibasis}).

It is important to realize that the gauge degrees of freedom in
Eq. (\ref{finallatticelagrangian})
are not independent from the fermionic degrees of freedom.
For example, neither $A_{\hat{\rm o}}$ nor the SU(2) factor of $W_{ij}$
possess a lattice version of the kinetic energy. Their presence is simply
a device to project the Fock space ${\cal F}$
which is generated cyclically from the
vacuum state $|0>_{\psi}$ defined by
\begin{equation}
\psi_{ia}\; |0>_{\psi}\;=\;0,\quad \forall i\;\in\;\Lambda,\quad a\;=\;1,2,
\end{equation}
onto the physical Hilbert space which is the tensorial product over
all sites $i$ of the vector spaces spanned by the color singlet states
$|0>_{\psi_i}$ and $b^{\dag}_i|0>_{\psi_i}$
(see \cite{particle-hole}).

\section{Mean-field theory and fluctuations around it}
\label{sec:s-RVBMF}

\subsection{The mean-field Ansatz}
\label{subsec:s-RVBAnsatz}

For the rest of the paper, we specialize to
a linear chain $\Lambda$ with nearest, $J^{\ }_1$,
and next-nearest, $J^{\ }_2$, neighbor antiferromagnetic couplings.
We try the translationally invariant Ansatz \cite{Wen 1991,Mudry 1994}
\begin{eqnarray}
&&\bar A^{\ }_{\hat{\rm o}it}\ =\ {1\over2}a^{1}_{\hat{\rm o}},
\nonumber\\
&&\bar W^{\   }_{ijt}\ =\
\cases{
-X\sigma^3& if $j=i+1$,\cr
-{\rm Re}\ E\ \sigma^1-{\rm Im}\ E\ \sigma^2& if $j=i+2$,\cr
0&otherwise.\cr
}
\label{mfansatz}
\end{eqnarray}
The mean-field spinon spectrum $\pm|\vec\xi_k|$ is then given by
\begin{eqnarray}
&&\xi^1_{k}=
-{a^1_{\hat{\rm o}}\over2}+{J_2\over2}\ {\rm Re}\ E\ \cos 2k,
\nonumber\\
&&\xi^2_{k}=
+{J_2\over2}\ {\rm Im}\ E\ \cos 2k,
\\
&&\xi^3_{k}=
+{J_1\over2}\ X\ \cos k.
\nonumber
\end{eqnarray}
The label $k$ denotes a reciprocal vector in the Brillouin zone $\Omega$.
Points of special significance in the Brillouin zone are the nodes
of the mean-field spectrum, i.e., those points $k^*\in\Omega$ with
\begin{equation}
|\vec\xi^{\ }_{k^*}|\ =\ 0.
\end{equation}
For example, if the only non-vanishing mean-field parameter is
$X$, then there are two nodes at $\pm {\pi\over2}$.
Nodes are absent from the mean-field
spectrum whenever the mean-field parameters $X$ and $E$ are simultaneously
non-vanishing \cite{pathological caseI}.
The values of the mean-field parameters
are determined by the saddle-point equations.

The saddle-point equations are
\begin{eqnarray}
&&
0\ =\
+{1\over |\Lambda|}\
\sum_{k\in\Omega}\
{\vec\xi^{\ }_k\over|\vec\xi^{\ }_k|},
\nonumber\\
&&
{\rm Re}\ E\ =\
+{1\over |\Lambda|}\
\sum_{k\in\Omega}\
{\xi^{1}_k\over|\vec\xi^{\ }_k|}\ \cos 2k,
\nonumber\\
&&
{\rm Im}\ E\ =\
-{1\over |\Lambda|}\
\sum_{k\in\Omega}\
{\xi^{2}_k\over|\vec\xi^{\ }_k|}\ \cos 2k,
\nonumber\\
&&
X\ =\
+{1\over |\Lambda|}\
\sum_{k\in\Omega}\
{\xi^{3}_k\over|\vec\xi^{\ }_k|}\ \cos k,
\end{eqnarray}
where $|\Lambda|$ is the (even)
number of sites in the chain and $\Omega$ is the
Brillouin Zone. They can be solved analytically in the two limits
${J^{\ }_2\over J^{\ }_1}=0$
and
${J^{\ }_1\over J^{\ }_2}=0$.
In the former case, the mean-field solution is
\begin{equation}
a^{1}_{\hat{\rm o}}\ =\ 0,\quad
E\ =\ 0,\quad
X\ =\ {2\over\pi}.
\end{equation}
The mean-field excitation spectrum is gapless at the two {\it discrete}
locations $\pm{\pi\over2}$ of the Brillouin zone.
In the latter case, the mean-field solution is
\begin{equation}
a^{1}_{\hat{\rm o}}\ =\ 0,\quad
{\rm Re}\ E\ =\ {2\over\pi},\quad
{\rm Im}\ E\ =\ 0,\quad
X\ =\ 0.
\end{equation}
The mean-field excitation spectrum is gapless at the four {\it discrete}
locations $\pm{\pi\over4}$ and $\pm{3\pi\over4}$ of the Brillouin zone.
The doubling of the nodes in the excitation spectrum when
${J^{\ }_1\over J^{\ }_2}=0$ is to be expected since in this limit
the Heisenberg model effectively decouples into two independent
Heisenberg models with antiferromagnetic nearest-neighbor interactions
$J^{\ }_2$ between the even and odd lattice sites, respectively. To sum up,
both limits yield the same mean-field excitation spectrum namely that
of a one dimensional tight-binding gas of fermions at {\it half-filling}
and describe one dimensional versions of the Baskaran-Zou-Anderson
(BZA) state \cite{Baskaran 1987}.

The qualitative features of the mean-field solutions for finite
${J^{\ }_2\over J^{\ }_1}$ can be understood in view of
the nature of the phase space in one dimension.
In one dimension, the gapless modes associated to nodes in the mean-field
spectrum have a dispersion relation which
can be linearized in the close vicinity of the nodes
since the nodes are isolated points in the Brillouin zone
\cite{pathological caseII}.
This situation is in contrast to the one in
two spatial dimensions where gapless modes
can have a dispersion relation which is quadratic in reciprocal space
(as is the case with the BZA state for which the nodes form lines)
or linear
(as is the case with the flux
state for which the nodes are isolated points \cite{Affleck 1988}).
Another consequence of the restrictive nature of phase space is that
there is no genuine flux phase since it is not possible to inclose flux
{\it locally} in one spatial dimension.

This observation on the nature of the gapless modes in one spatial dimension,
allows us to understand qualitatively the mean-field solution for finite
values of ${J^{\ }_2\over J^{\ }_1}$. Without loss of generality, we only
need to consider the effect of an arbitrary small perturbation of the limit
${J^{\ }_2\over J^{\ }_1}=0$. The unperturbed mean-field spectrum is
\begin{equation}
\omega^{\ }_k\ =\
-{J^{\ }_1 X\over2}\ \cos k
,\quad
-\pi\leq k\leq +\pi
,\qquad
X\ =\ {2\over\pi}.
\end{equation}
Let us see first whether an infinitesimal value of
${J^{\ }_2\over J^{\ }_1}$
can induce a non-vanishing value for ${\rm Im}\ E$ alone. In other words,
is there a solution to the mean-field equation
\begin{equation}
|X|\ =\
-\
{J^{\ }_2\over J^{\ }_1}
\int^{+\pi}_{-\pi}\ {dk\over2\pi}\
{
\cos^2(2k)
\over
\sqrt
{
\cos^2 k
\ +\
\left(
{J^{\ }_2\over J^{\ }_1}
{{\rm Im}\ E\over X}
\right)^2
\cos^2 (2k)
}
}.
\end{equation}
The answer is negative due to the minus sign on the right-hand side.
However, one immediately sees that the mean-field equation
\begin{eqnarray}
&&
\label{mfree}
|X|\ =\
{J^{\ }_2\over J^{\ }_1}
\int^{+\pi}_{-\pi}\ {dk\over2\pi}\
{
\left(
-
{a^1_{\hat{\rm o}}\over J^{\ }_2{\rm Re}\ E}
\ +\
\cos 2k
\right)
\cos 2k
\over
\sqrt
{
\cos^2 k
\ +\
\left(
{a^1_{\hat{\rm o}}\over J^{\ }_1X}
\ -\
{J^{\ }_2\over J^{\ }_1}
{{\rm Re}\ E\over X}
\cos 2k
\right)^2
}
}
\end{eqnarray}
can be approximately solved, since the integral on the right-hand side
is dominated by the two contributions in the range
$\pm{\pi\over2}-\varepsilon\leq k\leq\pm{\pi\over2}+\varepsilon$,
to yield
\begin{eqnarray}
&&
{
2\pi|X|
\over
{a^1_{\hat{\rm o}}\over J^{\ }_1{\rm Re}\ E}
\ +\
{J^{\ }_2\over J^{\ }_1}
}\ \approx
2
\ln
\left[
{
\sqrt
{
1\ +\
{1\over\varepsilon^2}
\left(
{a^1_{\hat{\rm o}}\over J^{\ }_1X}
\ +\
{J^{\ }_2\over J^{\ }_1}
{{\rm Re}\ E\over X}
\right)^2
}
\ +\
1
\over
\sqrt
{
1\ +\
{1\over\varepsilon^2}
\left(
{a^1_{\hat{\rm o}}\over J^{\ }_1X}
\ +\
{J^{\ }_2\over J^{\ }_1}
{{\rm Re}\ E\over X}
\right)^2
}
\ -\
1
}
\right].
\end{eqnarray}
One verifies that
Eq.\ (\ref{mfree})
has the solution
\begin{equation}
a^{1}_{\hat{\rm o}}\ =\ 0
,\quad
\left({\rm Re}\ E\over X\right)^2
\ \approx\
\varepsilon^2
\left(4{J^{\ }_2\over J^{\ }_1}\right)^{-1}
{\rm exp}\ \left(-{\pi|X|\over{J^{\ }_2\over J^{\ }_1}}\right),
\label{bcslikesolution}
\end{equation}
for any infinitesimal value of ${J^{\ }_2\over J^{\ }_1}$.

We thus see that the linearity of the unperturbed mean-field spectrum
in the vicinity of the nodes allows for the existence of simultaneous
non-vanishing values of the mean-field parameters $X$ and
${\rm Re}\ E$ for any infinitesimal ${J^{\ }_2\over J^{\ }_1}$.
Conversely, the same analysis holds infinitesimally close to the limit
${J^{\ }_1\over J^{\ }_2}=0$. It is natural to extrapolate that the
simultaneous condensation of ${\rm Re}\ E$ and $X$ takes place at the
mean-field level for any finite ${J^{\ }_2\over J^{\ }_1}$.
We have solved numerically the saddle-point equations for
$0\leq {J^{\ }_2\over J^{\ }_1}\leq 10$ in the thermodynamic limit.
The numerical solution confirms that ${\rm Re}\ E$ and $X$
approach monotonically ${2\over\pi}$ and 0, respectively,
as ${J^{\ }_2\over J^{\ }_1}$ is increased from 0 to 10.
The mean-field parameter $|a^{1}_{\hat{\rm o}}|$
reaches a maxima around ${J^{\ }_2\over J^{\ }_1}=1$
and quickly decreases. The mean-field spectrum has a gap within the
numerical precision for any finite amount of frustration
${J^{\ }_2\over J^{\ }_1}$.
The gap only closes when the mean-field Ansatz
approaches the one dimensional BZA states.

The opening of a gap in the mean-field excitation spectrum of our spin liquid
for any amount of frustration is a dramatic signal of the failure of the
mean-field theory in view of the argument given by Haldane
\cite{Haldane 1982}
that the frustrated one dimensional
spin-${1\over2}$ antiferromagnet remains gapless for
$
0
\leq
{J^{\ }_2\over J^{\ }_1}
<
\left(J^{\ }_2\over J^{\ }_1\right)_c
=
{1\over6}
$
{}.
Our mean-field theory only predicts quantum criticality in
the absence of frustration. However, even in this limit the mean-field
theory is highly unreliable \cite{Mudry 1994}. Indeed, the constraint of
single occupancy which is needed to establish the equivalence
between the spinon model and the Heisenberg model
is not satisfied locally but only on average in the mean-field approximation.
Consequently, one should not believe the mean-field prediction for,
say, staggered spin-spin correlations.

We are going to construct below
an effective low energy theory which includes enough of the dynamics
ignored by the mean-field approximation so as to recover the quantum
criticality for {\it small frustration} $J_2\over J_1$
and insure equivalence with the
low energy sector of the Heisenberg model. The restoration
of criticality does not imply that the mean-field excitations just
above the mean-field gap have coalesced into a gapless branch
of ``dressed'' mean-field excitations once the fluctuations
around the mean-field have been included.
Rather, the mean-field single-particle
excitations have {\it completely} disappeared from
the spectrum and the gapless branch describes totally different excitations,
namely topological excitations (solitons).

\subsection{Low energy fluctuations around the mean-field Ansatz}
\label{subsec:contlimit}

To begin with, we need some notation.
Let $\Lambda_e$ and $\Lambda_o$ be the sublattices of even and odd sites,
respectively,
\begin{eqnarray}
&&
\Lambda_e\ =\ \{i\in\Lambda|i\ {\rm mod}\ 2=0\},
\nonumber\\
&&
\Lambda_o\ =\ \{i\in\Lambda|i\ {\rm mod}\ 2=1\}.
\end{eqnarray}
Define for any given even site $i\in\Lambda_e$
\begin{eqnarray}
&&
f^1_{i}\ =\ \psi^{\ }_{i},\quad
f^2_{i}\ =\ \psi^{\ }_{(i+1)},
\nonumber\\
&&
A^1_{i}\ =\ A^{\ }_{\hat{\rm o}i},\quad
A^2_{i}\ =\ A^{\ }_{\hat{\rm o}(i+1)},
\nonumber\\
&&
M^1_{i}\ =\ W^{\ }_{i(i+1)},\quad
M^2_{i}\ =\ W^{\ }_{(i+1)(i+2)},
\nonumber\\
&&
Q^1_{i}\ =\ W^{\ }_{i(i+2)},\quad
Q^2_{i}\ =\ W^{\ }_{(i+1)(i+3)},
\nonumber\\
&&
U^1_{i}\ =\ U^{\ }_{i},\quad
U^2_{i}\ =\ U^{\ }_{i+1}.
\label{relabellingI}
\end{eqnarray}
The Lagrangian of Eq. (\ref{finallatticelagrangian})
is now given by
\begin{eqnarray}
L=&&
\sum_{i\in\Lambda_e}\
\left[
f^{1\dag}_{it}\
({\rm i}\partial^{\ }_t-A^{1}_{it})
f^{1\   }_{it}+
f^{2\dag}_{it}\
({\rm i}\partial^{\ }_t-A^{2}_{it})
f^{2\   }_{it}
\right]
\nonumber\\
-&&
{J_1\over4}
\sum_{i\in\Lambda_e}
\left[
{1\over2}\ {\rm tr}\ \left(M^{1\   }_{it}M^{1\dag}_{it}\right)
+
{1\over2}\ {\rm tr}\ \left(M^{2\   }_{it}M^{2\dag}_{it}\right)
\right]
\nonumber\\
-&&
{J_1\over4}
\sum_{i\in\Lambda_e}
\left(
f^{1\dag}_{it}\ M^{1\   }_{it}\ f^{2\   }_{it}+
f^{2\dag}_{it}\ M^{2\   }_{it}\ f^{1\   }_{(i+2)t}+
{\rm H.c.}
\right)
\nonumber\\
-&&
{J_2\over4}
\sum_{i\in\Lambda_e}
\left[
{1\over2}\ {\rm tr}\ \left(Q^{1\   }_{it}Q^{1\dag}_{it}\right)
+
{1\over2}\ {\rm tr}\ \left(Q^{2\   }_{it}Q^{2\dag}_{it}\right)
\right]
\nonumber\\
-&&
{J_2\over4}
\sum_{i\in\Lambda_e}
\left(
f^{1\dag}_{it}\ Q^{1\   }_{it}\ f^{1\   }_{(i+2)t}+
f^{2\dag}_{it}\ Q^{2\   }_{it}\ f^{2\   }_{(i+2)t}+
{\rm H.c.}
\right).
\label{gaugetransformedlagrangian}
\end{eqnarray}

Consider now the gauge transformation
($\sigma^0$ is the two by two unit matrix)
\begin{equation}
\psi_i\rightarrow\ U_i\ \psi_i,\quad\forall i\in\Lambda,
\label{gaugetrsfkillsutwo}
\end{equation}
where
\begin{equation}
U_i\ =\
({\rm i})^{-i}
\cases
{
\sigma^3& if $i$ $\in$ $\Lambda_e$,\cr
\sigma^0& if $i$ $\in$ $\Lambda_o$.\cr
}
\label{BZAgauge}
\end{equation}
Under this gauge transformation
the mean-field Ansatz Eq. (\ref{mfansatz}) becomes
($i$ mod 4 = 0)
\begin{eqnarray}
&&
\bar A^1_i
\ \rightarrow\
-{1\over2}a^{1}_{\hat{\rm o}}\sigma^1,
\nonumber\\
&&
\bar A^2_i
\ \rightarrow\
+{1\over2}a^{1}_{\hat{\rm o}}\sigma^1,
\nonumber\\
&&
\bar M^1_i\ \rightarrow\
-{\rm i} X\ \sigma^0,
\nonumber\\
&&
\bar M^2_i\ \rightarrow\
-{\rm i} X\ \sigma^0,
\nonumber\\
&&
\bar Q^1_i\ \rightarrow\
(-{\rm Re}\ E\ \sigma^1-{\rm Im}\ E\ \sigma^2),
\nonumber\\
&&
\bar Q^2_i\ \rightarrow\
(+{\rm Re}\ E\ \sigma^1+{\rm Im}\ E\ \sigma^2),
\end{eqnarray}
In this gauge, the average
${1\over2}(\bar A^1_i+\bar A^2_i)$
vanishes while the difference
${1\over2}(\bar A^1_i-\bar A^2_i)$
does not. Similarly, the average
${1\over2}(\bar Q^1_i+\bar Q^2_i)$
vanishes while the difference
${1\over2}(\bar Q^1_i-\bar Q^2_i)$
does not.

To obtain a naive continuum limit, we define first
\begin{equation}
{\rm v_F}\ =\ {J_1X\bar\epsilon\over4},\quad
\bar\epsilon\ =\ 2\epsilon,\quad
X\ =\ {2\over\pi},
\end{equation}
where $\epsilon$ ($\bar\epsilon$)
is the lattice spacing on $\Lambda$ ($\Lambda_e$).
The sum over the lattice points $i\in\Lambda^{\ }_e$ is approximated
by an integral:
\begin{equation}
\sum_{i\in\Lambda_e}\ \bar\epsilon\ \rightarrow\int dx.
\end{equation}
The second step consists in identifying the slow or
smooth variables. For example, the fermionic site variable
$f^{\alpha}_i$ becomes a field $u^{\alpha }_x$
which is smooth over the scale $\bar\epsilon$, i.e.,
one assumes that
\begin{equation}
f^{\alpha}_{(i+2)a}=\sqrt{\bar\epsilon}\
\left[
u^{\alpha}_{ax}
\ +\
\bar\epsilon\ \partial^{\ }_x u^{\alpha}_{ax}
\ +\
{\cal O}(\bar\epsilon^2)
\right]
{}.
\label{smoothspinonI}
\end{equation}
The upper index $\alpha=1,2$ will be associated below with
the components of a Dirac spinor in two space-time dimensions.
The lower index $a=1,2$ is the SU(2) color index.
Finally, $x$ labels the continuous spatial co-ordinate.
In the sector implementing the local constraint of single occupancy,
one assumes the smooth variables to be
\begin{eqnarray}
&&
A^1_i\ =\ {\rm v_F}\ ({\cal A}^0_x\ +\ \bar\phi^0\ +\ \phi^0_x),
\\
&&
A^2_i\ =\ {\rm v_F}\ ({\cal A}^0_x\ -\ \bar\phi^0\ -\ \phi^0_x).
\end{eqnarray}
In the presence of frustration $J_2\over J_1$,
the uniform fluctuating field ${\cal A}^0_x$ does not
pick up an expectation value in our mean-field theory in contrast
to the staggered fluctuation $\phi^0_x$.
The use of the upper index is motivated below
where we show that ${\cal A}^0_x$
can be interpreted as the scalar component of a non-Abelian gauge field.
The vector component of the non-Abelian gauge field comes from the uniform
fluctuations of the nearest-neighbor links provided one uses the
{\it non-linear} parametrization
\begin{eqnarray}
&&
M^1_i\ =\
-{\rm i}X\left[1\ +\ \epsilon\ (\varrho^{\ }_x\ +\ \rho^{\ }_x)\right]
e^{-{\rm i}\epsilon({\cal A}^1_x+\phi^1_x)},
\\
&&
M^2_i\ =\
-{\rm i}X\left[1\ +\ \epsilon\ (\varrho^{\ }_x\ -\ \rho^{\ }_x)\right]
e^{-{\rm i}\epsilon({\cal A}^1_x-\phi^1_x)}.
\end{eqnarray}
Besides the uniform and staggered su(2) fluctuating fields
${\cal A}^1_x$
and
$\phi^1_x$,
respectively, there are uniform and staggered determinant fluctuations
$\varrho^{\ }_x$ and $\rho^{\ }_x$, respectively.
Finally, on the next-nearest neighbor links,
we will distinguish between
the color singlet fluctuations
$r^{\alpha}_x$, $\alpha=1,2$,
from the su(2) fluctuation
$R^{\alpha}_x$, $\alpha=1,2$,
by choosing the {\it linear} parametrization
\begin{eqnarray}
&&
Q^1_i\ =\
{\rm v_F}\
\left[-{\rm i}r^1_x\sigma^0\ +\ \bar R^1\ +\ \tilde R^1_x\right],
\\
&&
Q^2_i\ =\
{\rm v_F}\
\left[-{\rm i}r^2_x\sigma^0\ +\ \bar R^2\ +\ \tilde R^2_x\right].
\label{linearpara}
\end{eqnarray}
Only
$R^{\alpha}_x=\bar R^{\alpha}+\tilde R^{\alpha}_x$
picks up an expectation value $\bar R^{\alpha}$
in the presence of frustration at the mean-field level. Separating
the $r$'s from the $R$'s parallels separating the determinant
fluctuations from the su(2) fluctuations on the nearest-neighbor links.
Integration over the $r$'s and $\rho$'s will turn out to generate a
crucial interaction which induces {\it Umklapp processes}
\cite{Haldane 1980}, without which our continuum theory cannot capture
the departure from criticality (dimerization) as the frustration reaches
a critical value \cite{Haldane 1982}.

We also need to infer the gauge transformation law of the fluctuating
fields. We restrict ourself to SU(2) gauge transformation
which are uniform within the unit cell labelled by even sites:
\begin{equation}
U^1_{it}=U^2_{it}\equiv G^{\ }_{xt},
\quad\forall i\in\Lambda^{\ }_e.
\end{equation}
To the same order in the lattice spacing, the gauge
transformation law for the fluctuating fields is
\begin{eqnarray}
&&
{\cal A}_{\mu}\rightarrow
G{\cal A}_{\mu}G^{-1}+
({\rm i}\partial^{\ }_{\mu}G)G^{-1},\quad\mu=0,1,
\nonumber\\
&&
\phi_{\mu}\rightarrow
G\phi_{\mu}G^{-1}-
(1-\delta_{0\mu})\
({\rm i}\partial^{\ }_{\mu}G)G^{-1},\quad\mu=0,1,
\nonumber\\
&&
\tilde R^{\alpha}\rightarrow
G\tilde R^{\alpha}G^{-1},\quad\alpha=1,2,
\nonumber\\
&&
\varrho\rightarrow\varrho
,\quad
\rho\rightarrow\rho,
\nonumber\\
&&
r^{\alpha}\rightarrow r^{\alpha},\quad\alpha=1,2.
\label{restrictedgaugetrs}
\end{eqnarray}

Collecting terms of lowest order in $\bar\epsilon$
yields the effective Lagrangian density
\begin{eqnarray}
{\cal L}=&&
\sum_{\alpha=1,2}{\rm v_F}
\left[
u^{\alpha\dag}
(
{\rm i}{\partial^{\ }_t\over{\rm v_F}}-{\cal A}^0
)\
u^{\alpha\   }
+
u^{\alpha\dag}
(-1)^{\alpha}
\left(\bar\phi^0+\phi^0\right)\
u^{\alpha\   }
\right]
\nonumber\\
-&&\
{\rm v_F}
\left[
{X\over2\epsilon^2}+{X\over\epsilon}\varrho+
{X\over2}(\varrho^2+\rho^2)
\right]
\nonumber\\
+&&\
{\rm v_F}
\left[
u^{1\dag}
(
{\rm i}\partial^{\ }_1+{\cal A}^1
)\
u^{2\   }
+
u^{2\dag}
(
{\rm i}\partial^{\ }_1+{\cal A}^1
)\
u^{1\   }
+
u^{1\dag}
{\rm i}\rho\
u^{2\   }
-
u^{2\dag}
{\rm i}\rho\
u^{1\   }
\right]
\nonumber\\
-&&\
{\rm v_F}{J_2J_1X\over16}\sum_{\alpha=1,2}{1\over2}\ {\rm tr}\
\left[
R^{\alpha}
R^{\alpha\dag}
\right]
-
{\rm v_F}{J_2\over4}\
\left[
u^{1\dag}
\
(R^1+R^{1\dag})\
u^{1    }
+
u^{2\dag}
\
(R^2+R^{2\dag})\
u^{2    }
\right]
\nonumber\\
-&&\
{\rm v_F}{J_2J_1X\over16}\sum_{\alpha=1,2}\
\left(
r^{\alpha}
\right)^2
+
{\rm v_F}{J_2\over4}
\left[
u^{1\dag}_{x}\
{\rm i} r^1_x\
u^{1    }_{x+\bar\epsilon}
+
u^{2\dag}_{x}\
{\rm i}r^2_x\
u^{2    }_{x+\bar\epsilon}
\ +\ {\rm H.c.}
\right].
\label{semiclassicallagrangian}
\end{eqnarray}
We have displayed the fields spatial dependency whenever necessary
to indicate that the limiting procedure
$\bar\epsilon\rightarrow 0$ should be treated with special care.
Indeed, some terms appear to vanish due to the Pauli principle
if the continuum limit is taken without caution.
Notice that the uniform determinant fluctuation $\varrho$
and the su(2) staggered fluctuation $\phi^{\ }_1$ of $M^1$ and $M^2$
have decoupled from the other dynamical degrees of freedom
to lowest order in $\bar\epsilon$. We will ignore them completely
from now on. We will also ignore the irrelevant additive constants in
Eq. (\ref{semiclassicallagrangian}).

In view of the linearity of the
mean-field spectrum in the limit ${J^{\ }_2\over J^{\ }_1}=0$,
we should be able, in this limit, to recast
Eq. (\ref{semiclassicallagrangian}) in the form of
a relativistic theory in two
space-time dimensions. To stress this point,
we introduce a new set of Pauli matrices $\vec\tau$ and define
the gamma matrices
\begin{equation}
\gamma^0=\tau^2,\quad
\gamma^1=-{\rm i}\tau^3,\quad
\gamma^5=\tau^1,
\label{gammamatricesI}
\end{equation}
which satisfy the usual algebra
\begin{equation}
\{\gamma^{\mu},\gamma^{\nu}\}=2g^{\mu\nu}
,\quad
g^{00}=-g^{11}=1
,\quad
g^{01}=+g^{10}=0.
\end{equation}
We also use the notation
\begin{equation}
D_{\mu}\ =\ \partial^{\ }_{\mu}+{\rm i}{\cal A}_{\mu}
,\quad
\partial^{\ }_{\mu}\ \equiv\
(
{1\over {\rm v_F}}\
{\partial\over\partial{t}}\ ,\
{\partial\over\partial{x}}
)\ \equiv\
(\partial^{\ }_0,\partial^{\ }_1)
\end{equation}
for the covariant derivative.
We can then rewrite Eq. (\ref{semiclassicallagrangian})
as
\begin{equation}
{\cal L}\ =\
{\cal L}^{\ }_0
\ +\
{\cal L}^{\ }_1
\ +\
{\cal L}^{\ }_2
\ +\
{\cal L}^{\bar\epsilon}_2
{}.
\label{semiclassicallagrangiandensity}
\end{equation}
We have separated the contributions
${\cal L}^{\ }_0$
and
${\cal L}^{\ }_1$
which are invariant under proper
Lorentz transformation in two space-time dimensions
and are given by
\begin{equation}
{\cal L}^{\ }_0\ =\
{\rm v_F}\
\bar u^{\   }\
{\rm i}\gamma^{\mu}\ D_{\mu}\
u^{\   },
\label{semiclassicallagrangiandensityO}
\end{equation}
and
\begin{equation}
{\cal L}^{\ }_1\ =\
-{\rm v_F}\
\left(
\bar u^{\   }\
{\rm i}\gamma^5\ \phi_0\
u^{\   }
\ +\
\bar u^{\   }
u^{\   }\
\rho
\ +\
{X\over2}\rho^2
\right),
\label{semiclassicallagrangiandensityI}
\end{equation}
respectively, from the contribution
\begin{eqnarray}
{\cal L}^{\ }_2\ =&&\
-
{\rm v_F}\
\bar u^{\   }\
{\rm i}\gamma^5\
\bar\phi^{\ }_0\
u^{\    }
\nonumber\\
-&&
{\rm v_F}{J_2J_1X\over16}\sum_{\alpha=1,2}{1\over2}\ {\rm tr}\
\left[
R^{\alpha}R^{\alpha\dag}
\right]
\nonumber\\
-&&\
{\rm v_F}{J_2\over4}\
\bar u^{\   }\
\gamma^0\
{
(
R^1+R^{1\dag}+
R^2+R^{2\dag}
)
\over2}\
u^{\    }
\nonumber\\
-&&\
{\rm v_F}{J_2\over4}\
\bar u^{\   }\
{\rm i}\gamma^5\
{
(
R^1+R^{1\dag}-
R^2-R^{2\dag}
)
\over2}\
u^{\    },
\label{semiclassicallagrangiandensityII}
\end{eqnarray}
and the contribution
\begin{eqnarray}
{\cal L}^{\bar\epsilon}_2\ =&&
-
{\rm v_F}{J_2J_1X\over16}\sum_{\alpha=1,2}\ \left(r^{\alpha}\right)^2
\nonumber\\
+&&\
{\rm i}
{\rm v_F}{J_2\over4}\
\sum_{\alpha=1,2}
\left(
u^{\alpha\dag}_x u^{\alpha}_{x+\bar\epsilon}\ r^{\alpha}_x
\ -\
{\rm H.c.}
\right)
,
\label{semiclassicallagrangiandensityIII}
\end{eqnarray}
which both describe the interactions due to the frustration $J_2\over J_1$.
In the absence of fluctuations,
the low energy sector of the mean-field theory is correctly described by
\begin{equation}
\bar{\cal L}\ =\
{\rm v_F}
\left[
\bar u^{\   }\
{\rm i}\gamma^{\mu}\ \partial^{\ }_{\mu}\
u^{\   }
\ +\
\bar u^{\   }\
{\rm i}\gamma^5\
\left(
{a^1_{\hat{\rm o}}\over2}
\ +\
{J_2\over2}\
{\rm Re}\ E\
\right)
\ \sigma^1\
u^{\    }
\right].
\label{meanfieldcriticallag}
\end{equation}

The contribution
Eq. (\ref{semiclassicallagrangiandensityII})
to the low energy effective theory
does not possess a full relativistic invariance.
On the other hand, we know that the Heisenberg model
in the limit of small frustration $J_2\over J_1$
is in the quantum critical regime
and therefore should have its low energy sector
described by a field theory characterized by a gapless spectrum
and relativistic invariance \cite{Haldane 1982,Affleck 1987}.
That our field theory fails to do so on both account
is an artifact of the mean-field theory.
We are now going to show that the quantum theory constructed from
Eq. (\ref{semiclassicallagrangiandensity})
in the limit of sufficiently small frustration
is equivalent to a relativistic quantum field theory with no mass gap.
In other words, we are first going to show that the quantum fluctuations
restore quantum criticality by removing any mean-field gap associated to
the condensation of $\phi^{\ }_0$ and the $R$'s. We then show that
the spinon interactions induced by the fluctuations $\rho$ and $r^{\alpha}$
are {\it irrelevant} for small enough frustration $J_2\over J_1$.

\subsection{Quantum fluctuations around the mean-field Ansatz}
\label{subsec:Quantumfluctuations}

The quantum theory for the fluctuations around the mean-field Ansatz
is constructed from the partition function
\begin{equation}
Z\ =\
\int {\cal D}\mu^{\ }_b
\int {\cal D}\mu^{\ }_f
\
e^{+{\rm i}\int dt\ {\cal L}},
\label{QuantumtheoryforfluctuationsI}
\end{equation}
where the bosonic and fermionic integration measures are
\begin{eqnarray}
&&
{\cal D}\mu^{\ }_b\ =\
{\cal D}\left[r^1,r^2\right]\
{\cal D}\left[R^1,R^2\right]\
{\cal D}\left[\phi^{\ }_0\right]\
{
{\cal D}\left[{\cal A}^{\ }_{\mu}\right]\
\over
{\cal V}^{-1}\
}
{\cal D}\left[\rho\right],
\nonumber\\
&&
{\cal D}\mu^{\ }_f\ =\
{\cal D}\left[\bar u,u\right],
\end{eqnarray}
respectively, and ${\cal L}$ is the Lagrangian density of
Eq. (\ref{semiclassicallagrangiandensity}).
The measure for the fields $\phi^{\ }_0$ and ${\cal A}^{\ }_{\mu}$
is the measure for the Lie algebra su(2). The measure
for the fields $r^{1(2)}$  and $R^{1(2)}$
is the product of the measure for real scalar fields
with the su(2) measure.
Finally, the measure for $\rho$ is the measure for real scalar fields.
The factor ${\cal V}^{-1}$ serves to remind us that the gauge for the
${\cal A}^{\ }_{\mu}$ has to be fixed.

We now show the important result that the
fields $R^{1(2)}$ decouple from all other fields
in the partition function
Eq. (\ref{QuantumtheoryforfluctuationsI}).
The $R$'s belong to the Lie algebra su(2) by construction.
Consequently, the Hermitean linear combinations
\begin{eqnarray}
&&
B^{\ }_+\ \equiv\
R^{1    }
\ +\
R^{1\dag}
\ +\
R^{2    }
\ +\
R^{2\dag}
\ =\
B^{\dag}_+,
\nonumber\\
&&
B^{\ }_-\ \equiv\
R^{1    }
\ +\
R^{1\dag}
\ -\
R^{2    }
\ -\
R^{2\dag}
\ =\
B^{\dag}_-,
\end{eqnarray}
also belong to the Lie algebra su(2).
If we choose to integrate over the fields ${\cal A}^{\ }_0$
and $\phi^{\ }_0$ before integrating over the $R$'s, then
\begin{equation}
{\cal A}^{\ }_0\rightarrow
{\cal A}^{\ }_0 -
{J_2\over8}B^{\ }_+,\quad
\phi^{\ }_0\rightarrow
\phi^{\ }_0 -
\bar\phi^{\ }_0-{J_2\over8}B_-,
\label{shiftintegrationvar}
\end{equation}
is a well defined shift of integration variable which
decouples the $R$'s from the other fields and eliminate any mean-field gap.

Quantum fluctuations of the {\it uniform} and {\it staggered}
components of the Lagrange multipliers which enforce the constraint of
single occupancy, thus remove the mean-field gap due to the condensation
of $\phi^{\ }_0$ and the $R$'s for any amount of frustration $J_2\over J_1$
and  restore the full relativistic invariance of
${\cal L}_0+{\cal L}_1+{\cal L}_2$.
We want to see how these and the remaining quantum fluctuations
of $\rho$ and ${\cal A}^{\ }_1$ affect physical observables like
the spin-spin correlation functions. We will treat the interaction
${\cal L}^{\bar\epsilon}_2$ separately.
To this end and
without loss of generality, we will only include the dynamics
contained in the partition function
\begin{eqnarray}
&&
Z\ =\
\int
{\cal D}
\left[
\phi^{\ }_0
\right]
\int
{\cal D}
\left[
{\cal A}^{\ }_{\mu}
\right]
\int
{\cal D}
\left[
\bar u,u
\right]
\
e^{+{\rm i}\int dt\ \left({\cal L}^{\ }_0+{\cal L}'_1\right)}
,
\label{QuantumtheoryforfluctuationsII}
\\
&&
{\cal L}^{\ }_0\ =\
\bar u^{\   }\
{\rm i}\gamma^{\mu}\
\left(
\partial^{\ }_{\mu}
\ +\
{\rm i}{\cal A}^{\ }_{\mu}
\right)\
u^{\   },
\label{strongqcdtwo}
\\
&&
{\cal L}'_1\ =\
-
\bar u^{\   }\
{\rm i}\gamma^5\ \phi_0\
u^{\   }
\ +\
{1\over2X}\left(\bar u\ u\right)^2.
\label{perturbationtoqcd}
\end{eqnarray}
We have set ${\rm v_F}=1$ and performed the Gaussian integration over
$\rho$. The mean-field parameter $X$ is the solution of the saddle-point
equations in the limit ${J^{\ }_2\over J^{\ }_1}=0$.

If we neglect the quantum fluctuations of the bosonic fields, it
appears that the scaling dimensions of physical observables could
depend continuously on the value of $X$ as the action of
Eq. (\ref{QuantumtheoryforfluctuationsII})
then reduces to a variant of the Thirring model \cite{Thirring 1958}
in the presence of background fields.
This will turn out not to be the case as we show in the next section.

\section{Non-Abelian bosonization}
\label{sec:Nonabelianboso}

We have constructed a theory for the quantum fluctuations around the
mean-field Ansatz of Eq. (\ref{mfansatz}) given by
Eq. (\ref{QuantumtheoryforfluctuationsII}).
The action in Eq. (\ref{QuantumtheoryforfluctuationsII})
resembles the action for the SU(2) Thirring model except for the
presence of the gauge fields. It is known that the four fermion interaction
is a marginal operator which changes the anomalous dimensions
of the fermionic correlation functions in a {\it pure} fermionic theory
\cite{Fradkin 1991}.
Such an interaction is crucial to derive the correct anomalous dimensions of
the staggered magnetization in the Jordan-Wigner representation of the
Heisenberg chain for spin-${1\over2}$. However, in our case this interaction
has highly undesirable consequences since the change in the anomalous
dimensions is a function of the mean-field parameter $X$, and not of a
physical parameter like the anisotropy in the Jordan-Wigner approach.

In this section, we {\it first} consider the quantum theory with
action ${\cal L}^{\ }_0$. We show how the quantum gauge fluctuations
can be treated {\it exactly} using {\it non-Abelian bosonization}.
As a by product, the physical states can be constructed explicitly
and the anomalous dimensions of the uniform and staggered
magnetizations extracted from the quantum theory
with the action ${\cal L}^{\ }_0$ are the correct one.
We then show that the quantum theory with action ${\cal L}^{\ }_0$
is stable with respect to the perturbation ${\cal L}'_1$ in the sense
that the additional interactions are irrelevant {\it and} do not modify
the values of the anomalous dimensions for the uniform and staggered
magnetizations.

\subsection{Non-Abelian bosonization of the quantum critical theory}
\label{subsec:bosoQCD}

We want to construct the physical states and
calculate the anomalous dimensions of the uniform and
staggered magnetizations from
\begin{eqnarray}
&&
Z_0\ =\
\int
{\cal D}
\left[
{\cal A}^{\ }_{\mu}
\right]
\int
{\cal D}
\left[
\bar u,u
\right]
\
e^{+{\rm i}\int dt\ {\cal L}^{\ }_0}
,
\label{QCDpartitionfunction}
\\
&&
{\cal L}^{\ }_0\ =\
\bar u^{\   }\
{\rm i}\gamma^{\mu}\
\left(
\partial^{\ }_{\mu}
\ +\
{\rm i}{\cal A}^{\ }_{\mu}
\right)\
u^{\   }.
\label{QCDaction}
\end{eqnarray}
The partition function $Z_0$
describes the critical theory in the limit of small frustration and
in the absence of the perturbation ${\cal L}'_1$. It is equivalent to the
partition function for the infinitely
strong coupling limit of $QCD_2$  with su(2)
color gauge fields. The role of the su(2) gauge fields ${\cal A}^{\ }_{\mu}$
is to project the spinon Fock space onto the subspace of color singlet
states which is
defined by the constraint
\begin{equation}
\vec J^{\mu}\ \equiv\ \bar u\ \gamma^{\mu}\ {\vec\sigma\over2}\ u\ =\ 0.
\label{colorconstraint}
\end{equation}
The effect of the perturbations due to the
fluctuations of the staggered gauge field $\phi^{\ }_0$ and of the
current-current interaction will be studied in the next subsection.

Our strategy
is to establish an equivalence between
the partition function Eq. (\ref{QCDpartitionfunction})
and the partition function of
a non-interacting theory to be described below which is
explicitly {\it conformally invariant}
\cite{Karabali 1990,Tanii 1990,Itoi locarno}.
We can then borrow general results from two-dimensional conformal
field theory to show explicitly how all the single particle
mean-field excitations are {\it removed} from the singlet sector of the
Fock space. Our conformal field theory turns out to be a special example
of a {\it coset model}
(specifically a conformal field theory on the homogeneous
space U(2)/SU(2) \cite{GKO,Karabali 1990}). This allows us to
construct the physical states from the energy-momentum tensor of $Z_0$.
Finally, the correct anomalous dimensions for the uniform and staggered
magnetizations are recovered. This last result makes explicit the
SU(2) spin dynamical symmetry which is hidden in $Z_0$.

The details of the construction of the equivalent theory can be found
in appendix \ref{sec:decouplinggauge}.
The idea is to use the vector and axial
symmetry of the Lagrangian density Eq. (\ref{QCDaction})
together with the property (unique to two space-time dimensions) that
$\gamma^{\mu}\gamma^5=\epsilon^{\mu\nu}\gamma^{\ }_{\nu}$,
to decouple the gauge fields from the spinons
in the Lagrangian density. The resulting action $S_1$ describes
free Dirac spinons.
However, because $Z_0$ in
Eq. (\ref{QCDpartitionfunction}) only shares the
vector gauge invariance of the Lagrangian density, a non-trivial
change in the fermionic measure under the transformation decoupling
the spinons from the gauge fields induces a Wess-Zumino-Witten
contribution $S_2$ \cite{Witten 1984,Polyakov}.
The Wess-Zumino-Witten field depends only on the gauge fields and
its kinetic energy is {\it negative definite}. Karabali and Schnitzer
\cite{Karabali 1990} have shown that the theory of free Dirac spinons
and a Wess-Zumino-Witten action with negative definite kinetic energy
is well defined provided a contribution $S_3$ needed to fix the gauge is also
accounted for. We will see that the role of the sectors $S_2$ and $S_3$
associated to the gauge fields is to remove unphysical states
from the spinon Fock space. To put it differently, the gauge fields
implement, {\it up to some scale} defined by
the effective low energy theory $Z_0$,
the {\it Gutzwiller projection}
necessary for the equivalence between
the parent lattice theory for the spinons
Eq. (\ref{firstlatticelagrangian}) and the Heisenberg model.

The quantum critical theory Eq. (\ref{QCDpartitionfunction})
is equivalent to the {\it conformally invariant} theory
\begin{eqnarray}
&&
Z_0\ =\
\int
{\cal D}\mu_b
\int
{\cal D}\mu_f
\
e^{+{\rm i}(S_1+S_2+S_3)},
\label{bosonizedaction}
\\
&&
S_1\ =\
\int\ {dx^+dx^-\over2}\
\big(
 u'^{\dag}_+\ {\rm i}\partial^{\ }_-\  u'_+
\ +\
 u'^{\dag}_-\ {\rm i}\partial^{\ }_+\  u'_-
\big),
\label{fermionsector}
\\
&&
S_2\ =\
-(1+2c_v)\ W^{\ }_-[\tilde G^{\ }_-],
\label{gaugesector}
\\
&&
S_3\ =\
\int\ {dx^+dx^-\over2}\
\Big[
{\rm tr}\ (\beta'^a_+\sigma^a\ {\rm i}\partial^{\ }_- \sigma^b\alpha'^b_+)
+
{\rm tr}\ (\beta'^a_-\sigma^a\ {\rm i}\partial^{\ }_+ \sigma^b\alpha'^b_-)
\Big].
\label{ghostsector}
\end{eqnarray}
The measures are defined in
Eqs. (\ref{bosonicmeasure}) and (\ref{fermionicmeasure})
and the relationships between the fields in
Eq. (\ref{bosonizedaction}) and the Dirac spinons
and gauge fields of
Eq. (\ref{QCDpartitionfunction}) are given by
Eqs.
(\ref{freerightspinon}),
(\ref{freeleftspinon}),
(\ref{wesszuminowittenfield}).
Eq. (\ref{bosonizedaction}) describes
three sectors: the free spinon sector ($S_1$), the Wess-Zumino-Witten
sector ($S_2$) with {\it negative level}
$
k=-(1+2\times2)=-5
$,
and the ghost sector ($S_3$).
The three sectors are individually conformally invariant and
do not interact with each other.
Hence, we can construct the entire Fock space by taking the tensor
product of the eigenstates of each individual sectors. However,
not all states constructed in this way are physical due to the existence
of states with negative definite norm coming from the Wess-Zumino-Witten
and ghost sectors or, equivalently, due to the constraint
Eq. (\ref{colorconstraint}).

We begin with the counting
of the physical degrees of freedom in
the partition function Eq. (\ref{bosonizedaction}).
The action of Eq. (\ref{bosonizedaction}) has a
traceless (conformal invariance) energy-momentum tensor.
Its light-cone components
$T^{0}_{\pm}$ are the sum of pairwise commuting energy-momentum tensors
corresponding to the spinon, gauge, and ghost sectors, respectively:
\begin{equation}
T^{0 }_{\pm}\ =\
T^{1 }_{\pm}\ +\
T^{2 }_{\pm}\ +\
T^{3 }_{\pm}.
\end{equation}
The algebras obeyed by the two copies $T^{n}_{\pm},\ n=0,1,2,3$
are the Virasoro algebras with central charges $C_n,\ n=0,1,2,3$.
The Virasoro central charges
(which count the degrees of freedom)
are related by
\begin{equation}
C_0\ =\
C_1\ +\
C_2\ +\
C_3.
\end{equation}
Since $S_1$ describes 2 free Dirac fermions, one has
$
C_1=2
$
\cite{Polyakovhouches,Karabali 1990}.
According to Knizhnik and Zamolodchikov \cite{Knizhnik 1984},
the Virasoro central
charge associated to a Wess-Zumino-Witten action of level $k$
is
\begin{equation}
C\ =\ {k\over k\ +\ c_v}\ {\rm dim\ su(2)}.
\end{equation}
Thus, in our case,
$
C_2=5.
$
Finally, the Virasoro central charge for the ghost sector is negative
and given by \cite{Polyakovhouches,Karabali 1990}
\begin{equation}
C_3\ =\ -2\ {\rm dim\ SU(2)}\ =-6.
\end{equation}
The central charge $C_0$ for the critical theory Eq.
(\ref{QCDpartitionfunction}) is therefore
\begin{equation}
C_0\ =\ 2\ +\ 5\ -\ 6\ =\ 1.
\end{equation}
Since the central charge $C_0$ counts the number of
physical degrees of freedom,
we see that the gauge and ghost sectors reduce the number $C_1=2$ of
mean-field degrees of freedom.

To stress this point more strongly, and to draw a connection with
coset conformal theories \cite{GKO} which will be very useful to us,
one rewrites the central charge as
\begin{equation}
C_0\ =\ 2\ -\ {1\over1+2}(2^2-1)\ +\ 0.
\label{arithmetic}
\end{equation}
The motivation for this arithmetic game is that the energy-momentum
tensors of a large class of quantum field theories
can be constructed from currents obeying Kac-Moody algebras
\cite{Sugawara 1968}. For example, consider the
two dimensional {\it quantum} currents with the $+$ chiral component
\begin{equation}
j^a_+\equiv{j^a_0+j^a_1\over2},
\quad a=1,\cdots,n,
\end{equation}
obeying the {\it equal-time} algebra
\begin{equation}
[j^a_+(x)\ ,\ j^b_+(y)]\ =\
{\rm i}f^{abc}j^c_+(x)\ \delta(x-y)
\ +\
k\ {{\rm i}\over2\pi}\
\delta^{ab}\
\delta'(x-y).
\label{kacmoody}
\end{equation}
The numbers $f^{abc}$ are the structure constants of SU($n$).
The term which is proportional to
the spatial derivative of the delta function is called the Schwinger term
\cite{Schwinger 1959}.
It arises from the quantum nature of the currents and
is multiplied by the integer number $k$.
The algebra defined by Eq. (\ref{kacmoody}) is called
a Kac-Moody algebra of level $k$. It was shown long time ago
\cite{Sugawara 1968} how to interpret the right-hand side of
\begin{equation}
T^{\ }_+\ =\ {2\pi\over k+c_v}
\ j^a_+\ j^a_+,
\label{sugawaraforsun}
\end{equation}
in order for $T^{\ }_+$ to describe the energy-momentum tensor of a local
quantum-field theory (see appendix \ref{sec:sugawara} for details).
Here, $c_v\delta^{aa'}= f^{abc} f^{a'bc}$ is the
quadratic Casimir invariant in the adjoint representation of SU($n$).
Assume now that to each central charge on the right-hand side of
Eq. (\ref{arithmetic}), there corresponds an energy-momentum tensor
which can be built from appropriate currents like in
Eq. (\ref{sugawaraforsun}).
The first number on the right-hand side would
coincide with the central charge of the energy-momentum tensor
$T^{{\rm U(2)}}_{\pm}$
which is built from
currents obeying a U(2) Kac-moody algebra of level 1.
The second (negative) number on the right-hand side would
coincide with the central charge of the energy-momentum tensor
$T^{{\rm SU(2)}}_{\pm}$
which is built from
currents obeying a SU(2) Kac-moody algebra of level 1.
The zero is meant to remind us that energy-momentum
tensors can have vanishing central charges \cite{QED}
and we allow for this possibility by denoting with
$T'_{\pm}$ the corresponding energy-momentum tensor.
Eq. (\ref{arithmetic})
is then very suggestive of the decomposition
\begin{equation}
T^{0}_{\pm}\ =\
T^{{\rm U(2)}}_{\pm}
\ -\
T^{{\rm SU(2)}}_{\pm}
\ +\
T'_{\pm}
\ \cong\
T^{{\rm U(2)/SU(2)}}_{\pm}
\ +\
T'_{\pm}.
\label{cosetstructureenergymomentum}
\end{equation}

In our problem, we have the
U(1) color singlet current
Eq. (\ref{singletNoether})
and the SU(2) color current
Eq. (\ref{colorNoether}) at disposal.
We show in appendix \ref{sec:sugawara} that the color singlet current
obeys an Abelian Kac-Moody algebra for two fermion species and that the
color SU(2) current obeys
a level one SU(2) Kac-Moody algebra so that $T^{1}_{\pm}$ can
be identified with $T^{{\rm U(2)}}_{\pm}$ and $T^{{\rm SU(2)}}_{\pm}$
can indeed be constructed. The construction of $T'_{\pm}$
has been done by Karabali and Schnitzer and it involves
the color and ghost currents. They have investigated
the nature of the second
``equality'' sign of Eq. (\ref{cosetstructureenergymomentum}).
The issue is delicate
since the theory defined by $T^0_{\pm}$ does not have a positive definite
metric \cite{Karabali 1990}.
It is sufficient here to interpret
Eq. (\ref{cosetstructureenergymomentum})
as the removal of all unphysical states induced by the spinon representation.
The physical Hilbert space is to be
constructed from the unitary representation of the Kac-Moody algebra
and associated Virasoro algebra generated by the color singlet currents
Eq. (\ref{singletNoether}). Details can be found in the work of
Karabali and Schnitzer but, for our purpose, the important point is that
the states of the physical Hilbert space defined by $Z_0$
are (vector) gauge singlets. In other words,
the mean-field one-particle excitations
of Eq. (\ref{meanfieldcriticallag}) have been entirely
projected out of the physical spectrum.

The conformal invariance of Eq. (\ref{bosonizedaction}) implies that
for each components of the uniform and staggered magnetizations
there exists two conformal weights \cite{Belavin 1984}. To see this, recall
that pairs of conformal weights determine the transformation law
of the {\it primary fields} under conformal transformations.
{}From the knowledge of the conformal weights one easily extracts the scaling
dimensions (anomalous dimensions) of the primary fields.
So, if we can show that the magnetizations are products of one primary field
from each different sectors, and hence primary fields themselves,
one can calculate their conformal weights and their scaling dimensions.
The primary fields of Eq. (\ref{bosonizedaction}) are the Dirac spinons $u'$,
the Wess-Zumino-Witten fields $\tilde G^{\ }_-$, and the ghosts
$\beta',\alpha'$.
By inspection of Eq. (\ref{uniformmagnetization})
\begin{equation}
\vec M^{\ }_{+i}\ =\
{1\over2}
\left(
\vec S^{\ }_i
+
\vec S^{\ }_{i+1}
\right)
\ \propto\
{1\over2}
\left(
\vec {\cal J}^{\ }_+
+
\vec {\cal J}^{\ }_-
\right),
\end{equation}
for the uniform magnetization $\vec M^{\ }_{+i}$
and by inspection of Eq. (\ref{staggeredmagnetization})
\begin{equation}
\vec M^{\ }_{-i}\ =\
{1\over2}
\left(
\vec S^{\ }_i
-
\vec S^{\ }_{i+1}
\right)
\ \propto\
{1\over2}
\left(
\vec {\cal K}^{\ }_{+-}
+
\vec {\cal K}^{\ }_{-+}
\right)
\end{equation}
for the staggered magnetization
$\vec M^{\ }_{-i}$, it is apparent that the uniform
magnetization is invariant under all chiral transformations of
Eq. (\ref{chiralgaugetsfI},\ref{chiralgaugetsfII})
while the staggered magnetization is only invariant under the diagonal
subgroup of vector gauge transformations. Hence, the sequence of
chiral transformations which decouples the spinons from the gauge fields
results in
\begin{equation}
\vec {\cal J}^{\ }_{\pm}\
\ =\
{1\over2}
\pmatrix
{
-(u'^{\dag}_{{\pm}1}u'^{\dag}_{{\pm}2}+u'^{\   }_{{\pm}2}u'^{\   }_{{\pm}1})\cr
+{\rm i}(u'^{\dag}_{{\pm}1}u'^{\dag}_{{\pm}2}
-u'_{{\pm}2}u'_{{\pm}1})\cr
+(u'^{\dag}_{{\pm}1}u'_{{\pm}1}+u'^{\dag}_{{\pm}2}u'_{{\pm}2})\cr
},
\end{equation}
and
\begin{eqnarray}
\vec {\cal K}^{\ }_{+-}
\ =&&\
{1\over2}
\pmatrix
{
-
(
u'^{\dag}_{+b}\tilde G^{* }_{-1b}
u'^{\dag}_{-2}
+
u'_{-2}
u'_{+b}\tilde G^{\ }_{-1b}
)
\cr
+{\rm i}
(
u'^{\dag}_{+b}\tilde G^{* }_{-1b}
u'^{\dag}_{-2}
-
u'_{-2}
u'_{+b}\tilde G^{\ }_{-1b}
)
\cr
+
(
u'^{\dag}_{+b}\tilde G^{* }_{-1b}
u'_{-1}
+
u'^{\dag}_{+b}\tilde G^{\ }_{-2b}
u'_{-2}
)
}.
\end{eqnarray}

The magnetizations are now solely expressed in terms of the primary fields
$u'$ and $\tilde G^{\ }_-$ of the conformal field theory
Eq. (\ref{bosonizedaction})
\cite{Itoi locarno}.
One can therefore associate two conformal
weights to each components of the magnetizations. We use a vector notation
for the conformal weights:
$
\left(
\vec\Delta^{\ }_+
,
\vec\Delta^{\ }_-
\right)
\left[\vec{\cal J}(\vec{\cal K})\right]
$.
In turn, the scaling dimensions of
the magnetizations are the sum of the conformal weights
in the $+$ and $-$ chiral sectors:
\begin{equation}
\vec\Delta  \left[\vec{\cal J}(\vec{\cal K})\right]
=
\vec\Delta_+\left[\vec{\cal J}(\vec{\cal K})\right]
+
\vec\Delta_-\left[\vec{\cal J}(\vec{\cal K})\right].
\end{equation}
Using Eq. (\ref{primaryconformalweight})
of appendix \ref{sec:decouplinggauge}, we find that
the mean-field prediction for the scaling dimensions of the
uniform magnetization is unchanged by the quantum fluctuations of the
gauge fields and are
\begin{equation}
\vec\Delta\left(\vec{\cal J}\right)
\ =\
\pmatrix
{
1\cr
1\cr
1\cr
}.
\label{dimensionuniformmag}
\end{equation}
The mean-field prediction for the scaling dimensions of the
staggered magnetization are, however,
changed by the quantum fluctuations of the
gauge fields. The gauge fields fluctuations effectively {\it reduce}
the mean-field predictions for the scaling dimensions to yield
\begin{equation}
\vec\Delta\left(\vec{\cal K}\right)
\ =\
\pmatrix
{
{1\over2}\cr
{1\over2}\cr
{1\over2}\cr
}.
\label{dimensionstaggeredmag}
\end{equation}

According
to Eq. (\ref{dimensionuniformmag}) (\ref{dimensionstaggeredmag}),
all three components of the vector for the scaling dimensions of the
uniform (staggered) magnetization are the same
as required by the spin SU(2) invariance.
They, moreover, agree
with the scaling dimensions derived from the Jordan-Wigner representation
of the Heisenberg chain \cite{Fradkin 1991}.
The isotropy is not surprising in view of the fact
that our mean-field Ansatz respects the spin symmetry. However, the
mean-field theory overestimates the scaling dimension of the staggered
magnetization. The quantum gauge fluctuations are needed to restore the correct
scaling dimension of the staggered magnetization.

The scaling dimension of the uniform magnetization is left
unchanged by the quantum gauge fluctuations as the uniform magnetization
generates a continuous symmetry, namely the spin SU(2) symmetry.
It is tempting to identify the energy-momentum tensor
$T^{{\rm U(2)/SU(2)}}$ in Eq. (\ref{cosetstructureenergymomentum})
with the energy-momentum tensor constructed from the currents
$\vec{\cal J}^{\ }_{\pm}$
through the Sugawara construction \cite{Sugawara 1968}.
This is done explicitly in appendix \ref{sec:sugawara}
where we also show that the currents $\vec{\cal J}^{\ }_{\pm}$
satisfy a level one Kac-Moody algebra. Hence, the spin SU(2) symmetry
of the Heisenberg chain appears as a {\it dynamical symmetry}
of the critical theory Eq. (\ref{QCDpartitionfunction})
and we recover the results of Affleck
and Haldane on a (non-Abelian) bosonization scheme of the Heisenberg chain
which makes explicit the spin symmetry
\cite{Affleck 1987}.

Having shown how the quantum gauge fluctuations restore quantum
criticality and the correct scaling dimensions of the mean-field
theory for a one dimensional spin liquid,
we now turn to the effect of the perturbation ${\cal L}'_1$
on the conformal field theory Eq. (\ref{bosonizedaction}).
We will show in the next subsection that the constraints imposed
by the quantum fluctuations of the gauge fields {\it and} of
$\phi^{\ }_0$ guaranty all the results derived thus far even in
the presence of the perturbation ${\cal L}'_1$.

\subsection{Perturbations of the critical theory}
\label{subsec:perturbationqcd}

We want to see how the staggered fluctuations of the Lagrange multipliers
enforcing the constraint of one spinon per site and the staggered
fluctuations of the determinant on the nearest-neighbor links
modify our previous results. We are going to show that the scaling dimensions
are still given by the scaling dimensions
of the unperturbed conformal field theory Eq. (\ref{bosonizedaction})
{\it if and only if} the fluctuations of $\rho$ are always
treated together with the fluctuations of
$\phi^{\ }_0$.

We consider the partition function $Z$ given by
Eq. (\ref{QuantumtheoryforfluctuationsII}).
The perturbation ${\cal L}'_1$ has two effects. First, in addition to the
constraints
\begin{equation}
\vec J^{\mu}\ =\ \bar u\ \gamma^{\mu}\ {\vec\sigma\over2}\ u\ =\ 0,
\label{constraintsI}
\end{equation}
there are the constraints
\begin{equation}
\bar u\ {\rm i}\gamma^{\  }_5\ \vec\sigma\ u\ =\ 0,
\label{constraintsII}
\end{equation}
due to the quantum fluctuations of the staggered gauge fields $\phi^{\ }_0$.
Second, there is a quartic spinon interaction $(\bar u u)^2$
which can be rewritten
\begin{equation}
\left(
\bar u^{\ }
\
u^{\ }
\right)^2
\ =\
+
{1\over3}\
\left(
\bar u^{\ }
\
{\rm i}\gamma^{\ }_5\
\vec\sigma
\
u^{\ }
\right)^2
\ -\
{4\over3}\
\vec J^{\mu}\cdot\vec J^{\ }_{\mu}.
\end{equation}
This quartic interaction is caused by staggered fluctuations of
the determinant of the nearest-neighbor link $W^{\ }_{i(i+1)}$ variable.
Alone, it would change the scaling dimensions of the
staggered magnetization.
However, the quartic interaction vanishes identically on the states
annihilated by the left-hand sides of
Eq. (\ref{constraintsI})
and (\ref{constraintsII}).

To better understand the role of the constraints Eq. (\ref{constraintsII})
recall that the Lagrangian density ${\cal L}^{\ }_0$ of our critical theory
is invariant under local chiral transformations
Eq. (\ref{chiralgaugetsfI},\ref{chiralgaugetsfII})
and global chiral transformations
Eq. (\ref{chiralu1tsfI},\ref{chiralu1tsfII}).
The perturbation ${\cal L}'_1$ lowers
the symmetry of ${\cal L}^{\ }_0$ down to the vector subgroup
together with the {\it discrete} subgroup
\begin{equation}
u\rightarrow -{\rm i}e^{+{\rm i}{\pi\over2}\gamma^{\ }_5}\ u.
\label{discreteaxialsym}
\end{equation}
This is not surprising since the pure axial symmetry of ${\cal L}^{\ }_0$
has no counterpart in the rewriting of the Heisenberg model
as a lattice gauge theory.
The discrete chiral symmetry simply indicates that our
mean-field Ansatz does not break the translational invariance by one lattice
spacing as would be the case for an Ansatz with long range
antiferromagnetic order.
To see this last point, it is instructive to look at
the origin of the constraints
Eq. (\ref{constraintsII})
and
Eq. (\ref{constraintsI})
on the unit cell labelled by $i\in\Lambda^{\ }_e$.

The Fock space spanned by the two spinons
$s^{\ }_{i\sigma}$
and
$s^{\ }_{(i+1)\sigma}$
where $\sigma=\uparrow,\downarrow$,
is 16 dimensional.
It can be decomposed as the direct sum of the Hilbert spaces ${\cal H}^n$
with total spinon occupation number $n$ ranging from 0 to 4.
The physical subspace of the Fock space is the four dimensional
Hilbert space ${\cal H}^2_{phy}$ with one spinon per site.
It is to be distinguished from its complementary (with respect to
${\cal H}^2$) subspace ${\cal H}^2_{unphy}$
which has two spinons on either one of the two sites.

There are two operators which will characterize uniquely the physical
states of the Fock space if one requires the physical states
to be annihilated by  these operators, namely
\begin{equation}
\hat O^{3 }_+\ =\
\left(
s^{\dag}_i\ s^{\ }_i
\ -\
1
\right)
\ +\
\left(
s^{\dag}_{i+1}\ s^{\ }_{i+1}
\ -\
1
\right),
\end{equation}
and
\begin{equation}
\hat O^{3}_-\ =\
\left(
s^{\dag}_i\ s^{\ }_i
\ -\
1
\right)
\ -\
\left(
s^{\dag}_{i+1}\ s^{\ }_{i+1}
\ -\
1
\right).
\end{equation}
However, this choice is not unique since
\begin{equation}
\hat O^{\ }_+\ =\
s^{\dag}_{ i   \uparrow  }\ s^{\dag}_{ i   \downarrow}
\ +\
s^{\dag}_{(i+1)\uparrow  }\ s^{\dag}_{(i+1)\downarrow},
\end{equation}
and
\begin{equation}
\hat O^{\ }_-\ =\
s^{\dag}_{ i   \uparrow  }\ s^{\dag}_{ i   \downarrow}
\ -\
s^{\dag}_{(i+1)\uparrow  }\ s^{\dag}_{(i+1)\downarrow},
\end{equation}
or $\hat O^{\dag}_{\pm}$ perform the same task.

With the choice of Eq. (\ref{gammamatricesI}) for the gamma matrices
and by retracing all the steps
relating the Dirac spinons to the original spinons of
Eq. (\ref{spininspinonbasis}),
one can relate the local constraints in the continuum to constraints on
the states of the unit cell $i\in\Lambda^{\ }_e$. For example,
\begin{eqnarray}
\bar u\ \gamma^1\sigma^3\  u
\ &&=\
u^{\dag}\ \tau^1\sigma^3\  u
\nonumber\\
&&=\
+f^{1\dag}\ \sigma^3\ f^{2\ }
+f^{2\dag}\ \sigma^3\ f^{1\ }
\nonumber\\
&&\propto\
-{\rm i}
\left(
 \psi^{\dag}_{i  }\ \sigma^3\sigma^3\ \psi^{\   }_{i+1}
+\psi^{\dag}_{i+1}\ \sigma^3\sigma^3\ \psi^{\   }_{i  }
\right)
\nonumber\\
&&\propto\
-{\rm i}
\left(
 \psi^{\dag}_{i  }\ \psi^{\   }_{i+1}
+\psi^{\dag}_{i+1}\ \psi^{\   }_{i  }
\right)
\nonumber\\
&&=\
-{\rm i}
\left(
 s^{\dag}_{ i   \uparrow  }\ s^{\   }_{(i+1)\uparrow  }
+s^{\dag}_{ i   \downarrow}\ s^{\   }_{(i+1)\downarrow}
+{\rm H.c.}
\right),
\end{eqnarray}
tells us that the constraint on the space component of the color current
$\vec J^{\ }_1=0$
is related to spin currents in the unit cell $i\in\Lambda^{\ }_e$.
Similarly,
\begin{equation}
\bar u\ \gamma^0\vec\sigma\ u
\propto
\pmatrix{
-
\left[
\hat O^{\   }_-
\ +\
\hat O^{\dag}_-
\right]
\cr
+{\rm i}
\left[
\hat O^{\   }_-
\ -\
\hat O^{\dag}_-
\right]
\cr
\hat O^{3   }_+
}
\end{equation}
and
\begin{equation}
\bar u\ \gamma^1\vec\sigma\ u
\propto
\pmatrix{
-
\left[
\hat O^{\   }_+
\ +\
\hat O^{\dag}_+
\right]
\cr
+{\rm i}
\left[
\hat O^{\   }_+
\ -\
\hat O^{\dag}_+
\right]
\cr
\hat O^{3   }_-
}
\end{equation}
relate the original constraints and the two time-like constraints of
the continuum theory.

We see that the role of the staggered fluctuations $\phi^{\ }_0$
is to insure that there are as many spinons on either basis of the unit cell
$i\in\Lambda^{\ }_e$, whereas the role of the uniform scalar gauge
fluctuation is to insure that there is an average of two spinons per
unit cell. The lattice counterpart of the discrete symmetry
Eq. (\ref{discreteaxialsym}) is simply the exchange
\begin{equation}
c^{\ }_{ i    \uparrow  }
\ \rightarrow\
+c^{\ }_{(i+1)\uparrow  }
,
\qquad
c^{\ }_{ i    \downarrow}
\ \rightarrow\
-c^{\ }_{(i+1)\downarrow}
{}.
\end{equation}
In other words, multiplication of $u$ by $\gamma^5$
amounts to an interchange of the upper and lower components of $u$,
which on the lattice implies an interchange of even and odd sites.
There, is an additional flipping of the spins, due to the particle-hole
transformation Eq. (\ref{particleholetrs})
and a spin dependent sign change due to
the gauge transformation  Eq. (\ref{BZAgauge}).

In summary, our quantum critical theory correctly describes the low
energy sector of the Heisenberg chain for small frustration
although it only treats the
constraint of spinon single occupancy on average over the unit cell
$\Lambda^{\ }_e$. Beyond this microscopic scale, the constraint of
single occupancy is exactly satisfied. By rewriting the quantum critical
theory as a coset conformal theory, the dynamical spin SU(2) symmetry
manifests itself explicitly and contact is made with the non-Abelian
bosonization scheme of Affleck and Haldane for quantum spin chains
\cite{Affleck 1987}. All the one particle mean-field excitations,
the spinons, have been projected out of the physical Hilbert space.
The gapless modes carrying spin-${1\over2}$ are topological excitations
(solitons) which change the boundary conditions.
In two space-time dimensions,
it costs an infinite amount of energy to break the gauge singlet
bound states of spinons carrying integer spin quantum number and
deconfinement is not possible as a mechanism for
spin and charge separation.

\section{Dimerization}
\label{sec:Dimerization}

In the previous two sections, we have shown that the quantum theory
${\cal L}_0+{\cal L}'_1+{\cal L}_2$ in
Eq. (\ref{semiclassicallagrangiandensity})
is insensitive to {\it any} frustration ${J_2\over J_1}$.
However, we know from the exact ground state
and the low lying excitations of the Heisenberg chain
when ${J_2\over J_1}={1\over2}$
that criticality cannot hold for all values
of the frustration ${J_2\over J_1}$
\cite{j2/j1=0.5}.
Haldane \cite{Haldane 1982} has argued
that criticality of the system at ${J_2\over J_1}=0$
subsists up to a critical value
$\left({J_2\over J_1}\right)_c={1\over6}$.
The existence of a critical value for the frustration has been
confirmed numerically \cite{Tonegawa 1987}.
Haldane's analysis starts with the representation of the Heisenberg chain
in terms of Jordan-Wigner fermions (solitons). He
shows that, as the frustration ${J_2\over J_1}$ is switched on, an Umklapp
interaction which is {\it marginally irrelevant}
initially becomes relevant for a
finite value of the frustration and drives
the system into a massive phase characterized by long range dimer order.
First, we want to see if the
perturbation ${\cal L}^{\bar\epsilon}_2$,
Eq. (\ref{semiclassicallagrangiandensityIII}),
drives the system away from the level $k=1$ Wess-Zumino-Witten
fixed point and into a phase with dimer long range order.
Second, we derive the critical theory in the limit
${J_1\over J_2}=0$ and investigate what are the perturbations induced by
a small frustration ${J_1\over J_2}$ which can drive the system towards
dimerization. This issue is of relevance to the problem of two weakly
coupled Heisenberg chains.

\subsection{Relevant perturbation around ${J_2\over J_1}=0$}
\label{subsec:umklapp}

It is tempting, on the basis of Haldane's argument,
to believe that Umklapp processes for the {\it spinons}
are responsible for dimerization. However, one needs to be careful with
this analogy. Indeed, the relationship between Umklapp processes for the
Jordan-Wigner fermions,
which are the {\it gauge invariant}
spin-${1\over2}$
gapless modes of topological character
in the critical theory,
and Umklapp processes for the spinons is not obvious.

Umklapp processes for the spinons
are scattering events
in which a pair of excitation close
to the Fermi point $+k_{\rm F}=+{\pi\over2}$
is annihilated and a pair of
excitations close to the Fermi point $-k_{\rm F}$ is created or
{\it vice et versa}. Moreover,
all participants to this scattering event have the {\it same}
color quantum number.
Such processes are consistent with momentum conservation
at half-filling but explicitly {\it break}
the chiral symmetry of ${\cal L}_0$,
Eq. (\ref{semiclassicallagrangiandensityO}).
For example, in terms of the chiral components of
our Dirac spinors [see Eq. (\ref{weylspinon})],
two possible Umklapp processes result from the interactions
\begin{equation}
u^{* }_{-a x              }
u^{* }_{-a(x+\bar\epsilon)}
u^{\ }_{+a x              }
u^{\ }_{+a(x+\bar\epsilon)}
,\quad a=1\ or\ 2.
\end{equation}
Umklapp processes for the spinons
are {\it consistent} with local vector gauge invariance
since they all are induced by interactions of the form
\begin{eqnarray}
&&
\left(
u^{* }_{-a x              }\ \delta^{\ }_{ab}\ u^{ }_{+b x              }
\right)
\left(
u^{* }_{-c(x+\bar\epsilon)}\ \delta^{\ }_{cd}\ u^{ }_{+d(x+\bar\epsilon)}
\right),
\\
&&
\left(
u^{* }_{+a x              }\ \delta^{\ }_{ab}\ u^{ }_{-b x              }
\right)
\left(
u^{* }_{+c(x+\bar\epsilon)}\ \delta^{\ }_{cd}\ u^{ }_{-d(x+\bar\epsilon)}
\right),
\end{eqnarray}
or, equivalently, by $(\bar uu)^2$ and $(\bar u{\rm i}\gamma^5 u)^2$.
Notice that to obtain a representation of the Umklapp process
in a quantum field theory,
one needs to account for singularities associated with
multiplication of quantum fields at the same point,
e.g., with the procedure of point splitting
given in appendix \ref{sec:sugawara}.

The only gauge invariant fluctuating fields at our disposal are the
nearest neighbor staggered determinant fluctuations $\rho$ and
the next-nearest neighbor fluctuations $r^{\alpha}$, $\alpha=1,2$.
Integrating over $\rho$ and $r^{\alpha}$ induces Umklapp interactions
for the spinons with {\it coupling strength}
$
{\rm v_F}{1\over2X},
$
and
$
{\rm v_F}{1\over2X}\times4{J_2\over J_1},
$
respectively. What about their relative sign?

Instead of determining the relative sign from the field theory,
it is instructive to go back to the original pure spinon lattice theory,
Eq. (\ref{firstlatticelagrangian}).
Recall that the quartic interactions
originate from
\begin{eqnarray}
&&
\chi^{* }_{ij}\ =\
s^{* }_{i\uparrow  }\ s^{\ }_{j\uparrow  }
\ +\
s^{* }_{i\downarrow}\ s^{\ }_{j\downarrow}
,
\\
&&
\eta^{* }_{ij}\ =\
s^{* }_{i\uparrow  }\ s^{* }_{j\downarrow}
\ -\
s^{* }_{i\downarrow}\ s^{* }_{j\uparrow  }
,
\end{eqnarray}
or equivalently from
\begin{eqnarray}
&&
\chi^{* }_{ij}\ =\
\psi^{* }_{i1}\ \psi^{\ }_{j1}
\ -\
\psi^{* }_{j2}\ \psi^{\ }_{i2}
,\\
&&
\eta^{* }_{ij}\ =\
\psi^{* }_{i1}\ \psi^{\ }_{j2}
\ +\
\psi^{* }_{j1}\ \psi^{\ }_{i2}
,
\end{eqnarray}
where
$
\psi^{\ }_{i1}\ =\ s^{\ }_{i\uparrow  }
,\quad
\psi^{\ }_{i2}\ =\ s^{* }_{i\downarrow}
{}.
$
Umklapp processes originate solely from the
Affleck-Marston order parameter $\chi^{* }_{ij}$ since
\begin{eqnarray}
\chi^{* }_{ij}\ \chi^{\ }_{ij}\ =&&\
\psi^{* }_{i1}
\psi^{* }_{j1}
\psi^{\ }_{i1}
\psi^{\ }_{j1}
\ +\
\psi^{* }_{j2}
\psi^{* }_{i2}
\psi^{\ }_{j2}
\psi^{\ }_{i2}
\nonumber\\
+&&\
\psi^{* }_{i1}
\psi^{* }_{i2}
\psi^{\ }_{j1}
\psi^{\ }_{j2}
\ +\
\psi^{* }_{j2}
\psi^{* }_{j1}
\psi^{\ }_{i2}
\psi^{\ }_{i1}
,
\end{eqnarray}
whereas the color index of the annihilated pair of spinons always differ in
\begin{eqnarray}
\eta^{* }_{ij}\ \eta^{\ }_{ij}\ =&&\
\psi^{* }_{i1}
\psi^{* }_{j2}
\psi^{\ }_{i1}
\psi^{\ }_{j2}
\ +\
\psi^{* }_{i1}
\psi^{* }_{i2}
\psi^{\ }_{j1}
\psi^{\ }_{j2}
\nonumber\\
+&&\
\psi^{* }_{j1}
\psi^{* }_{j2}
\psi^{\ }_{i1}
\psi^{\ }_{i2}
\ +\
\psi^{* }_{j1}
\psi^{* }_{i2}
\psi^{\ }_{j1}
\psi^{\ }_{i2}
{}.
\end{eqnarray}
Thus, Umklapp processes for the spinons are caused by the interaction
$\chi^{* }_{ij}\chi^{\ }_{ij}$ through
\begin{eqnarray}
K^{\ }_{ij}\ =&&\
\psi^{* }_{i1}
\psi^{* }_{j1}
\psi^{\ }_{i1}
\psi^{\ }_{j1}
\ +\
\psi^{* }_{i2}
\psi^{* }_{j2}
\psi^{\ }_{i2}
\psi^{\ }_{j2}.
\end{eqnarray}
We need the continuum limit of $K^{\ }_{ij}$ when $j$ is either a
nearest or next-nearest neighbor of $i$.
To take the continuum limit, we use the variables
$f^1_i$ and $f^2_i$ defined in
Eqs. (\ref{relabellingI}),
perform the gauge transformation Eq. (\ref{gaugetrsfkillsutwo}),
and rewrite the interaction in terms of the chiral components
of the smooth fields $u^{\alpha}_{ax}$, $\alpha=1,2$, $a=1,2$.
In this way, we
extract the Umklapp terms
\begin{equation}
-
{\bar\epsilon^2\over4}\sum_{a=1,2}\
\left[
u^{* }_{-ax               }
u^{* }_{-a(x+    \epsilon)}
u^{\ }_{+ax               }
u^{\ }_{+a(x+    \epsilon)}
\ +\
(-\leftrightarrow+)
\right]
\end{equation}
from $K^{\ }_{i(i+1)}$ and $K^{\ }_{(i+1)(i+2)}$.
On the other hand, one extracts
\begin{equation}
+
{\bar\epsilon^2\over4}\sum_{a=1,2}\
\left[
u^{* }_{-ax               }
u^{* }_{-a(x+\bar\epsilon)}
u^{\ }_{+ax               }
u^{\ }_{+a(x+\bar\epsilon)}
\ +\
(-\leftrightarrow+)
\right]
\end{equation}
from $K^{\ }_{i(i+2)}$ and
$K^{\ }_{(i+1)(i+3)}$.
We obtain the important result that
a given Umklapp process
is induced by the interaction $\chi^{* }_{i(i+1)}\chi^{\ }_{i(i+1)}$
as well as by the interaction $\chi^{* }_{i(i+2)}\chi^{\ }_{i(i+2)}$
but with couplings of {\it opposite} sign. The magnitude of the
Umklapp coupling will thus be proportional to
$
\left|
1-4{J_2\over J_1}
\right|.
$

The instability of the critical theory ${\cal L}^{\ }_0$ towards
Umklapp processes for the spinons follows at once if we can show that
Umklapp processes can be induced by a {\it gauge invariant perturbation}
of ${\cal L}_0$
which is {\it marginally irrelevant} when ${J_2\over J_1}<{1\over4}$,
and is {\it marginally relevant} when ${J_2\over J_1}>{1\over4}$.

As we have shown in the Sec. \ref{subsec:bosoQCD},
the critical theory
described by ${\cal L}_0$ is equivalent to a level $k=1$ Wess-Zumino-Witten
theory constructed from the level one Kac-Moody SU(2) currents
$\vec{\cal J}^{\ }_{\pm}$. These currents are color singlets.
Their SU(2) algebra is related to the underlying
spin symmetry of the problem.
The only relevant perturbation to the level $k=1$ Wess-Zumino-Witten
theory which respects the diagonal chiral invariance {\it and} the discrete
chiral invariance Eq. (\ref{discreteaxialsym})
is $\vec{\cal J}^{\ }_+\cdot\vec{\cal J}^{\ }_-$
\cite{Knizhnik 1984,Affleck 1986}.
It turns out that this interaction is marginally relevant or irrelevant
depending on the sign of its coupling constant
\cite{Affleck 1987}.
In appendix \ref{sec:identities}, we relate
the Umklapp interaction for {\it spinons} to
$\vec{\cal J}^{\ }_+\cdot\vec{\cal J}^{\ }_-$
with the help of Eq. (\ref{umklappidentity}).
The crucial point is that both interactions
have the {\it same action} on the physical states,
since they only differ by the gauge invariant interactions
$
\left(\bar u\ {\rm i}\gamma^5\ u\right)^2
$
and
$
\vec J^{\ }_+\cdot\vec J^{\ }_-,
$
which are both constrained to annihilate the physical states of the theory.

In the absence of frustration ${J_2\over J_1}$, we have shown
in Sec.  \ref{subsec:perturbationqcd} that
Umklapp processes for the spinons
are irrelevant at the fixed point
corresponding to the level $k=1$ Wess-Zumino-Witten theory.
These Umklapp processes result from
the fluctuations of the staggered determinant $\rho$.
We know from the construction of the continuum limit that
the fluctuations $r^{\alpha}$, $\alpha=1,2,$ only induce
Umklapp processes
(${\cal L}^{\bar\epsilon}_2$ vanishes to lowest order in $\bar\epsilon$),
and since they come with a sign opposite to that due to the $\rho$'s
they are marginally relevant perturbations to the critical theory.
The competition between the
nearest and next-nearest Umklapp processes yield an instability
of the critical theory for
$\left({J_2\over J_1}\right)_c={1\over4}$.
For frustration larger than the critical one, the order parameter
$\rho$ develops spontaneously an expectation value corresponding
to the onset of dimerization. Note that our critical value for the
frustration differs numerically from Haldane's. This is to be
expected since this number is not universal but depends on the
short distance cutoff used.

We have thus recovered qualitatively the analysis of Haldane
\cite{Haldane 1982} by using a spinon representation of the
Heisenberg chain and starting from a mean-field theory around
a spin liquid. A {\it necessary} ingredient to this reconstruction
is to implement {\it exactly} the local constraint of single occupancy
in the spinon Fock space. Another {\it necessary}
ingredient is to treat the Affleck-Marston
and Anderson order parameters on an {\it equal} footing. Together, these
two ingredients amount to a {\it non-perturbative} treatment of the SU(2)
color symmetry.

\subsection{An effective field theory around the limit ${J_1\over J_2}=0$}
\label{subsec:twochains}

The frustrated spin-${1\over2}$ chain is solvable when
${J_1\over J_2}=0$, being equivalent to two
independent antiferromagnetic Heisenberg chains with nearest neighbor
interaction $J_2$. An interesting question is what is the effect
of the frustration ${J_1\over J_2}$, i.e., is it a relevant perturbation
or is it irrelevant up to some critical value
$\left({J_1\over J_2}\right)_c$.
It has been argued for the
closely related problem of the spin-${1\over2}$ ladder
that any interaction across the rung is relevant and
induces dimerization \cite{doublechain}.
We have found in Sec. \ref{subsec:s-RVBAnsatz}
that our mean-field theory predicts
a short-range RVB state in the presence of any infinitesimal frustration.
A short-range RVB state ($\bar\rho=0$)
certainly does not carry long range dimer order
($\bar\rho\not=0$).
However, it can be argued that it is unstable towards
dimerization. Here, we want to point out, on the basis
of the symmetry of a field theory for the fluctuations
around the short-range BZA spin liquid,
that other instabilities are present as well
(three besides the instability towards dimerization).
We also show that the mechanism restoring criticality for small
${J_2\over J_1}$ is not present in our field theory.

We begin by writing down the field theory at criticality. It is constructed
from two species of spinons $u$ and $v$ (one for each chains).
We only need to replace
Eq. (\ref{smoothspinonI}) by
\begin{eqnarray}
&&
f^{e\alpha}_{(i+4)a}\ =\
\sqrt{2\bar\epsilon}\
\left[
u^{\alpha}_{ax}
\ +\
2\bar\epsilon
\partial^{\ }_x
u^{\alpha}_{ax}
\ +\
{\cal O}(\bar\epsilon^2)
\right]
,
\\
&&
f^{o\alpha}_{(i+4)a}\ =\
\sqrt{2\bar\epsilon}\
\left[
v^{\alpha}_{ax}
\ +\
2\bar\epsilon
\partial^{\ }_x
v^{\alpha}_{ax}
\ +\
{\cal O}(\bar\epsilon^2)
\right]
,
\end{eqnarray}
where
$i\ {\rm mod}\ 4=0$,
and
$f^{e1}_{ia}=\psi^{\ }_{ia}$,
$f^{o1}_{ia}=\psi^{\ }_{(i+1)a}$,
$f^{e2}_{ia}=\psi^{\ }_{i(a+2)}$,
$f^{o2}_{ia}=\psi^{\ }_{i(a+3)}$,
$a=1,2$ being the color index.
Our critical theory depends on {\it twice} as many slow
variables as the single chain critical theory.
We borrow the notation from Sec. \ref{subsec:contlimit} for the
slow bosonic variables adding only an upper index $e$ and $o$
where necessary. The relevant kinetic scale is
${\rm v_F}={J_2{\rm Re}\ E\bar\epsilon\over2}$. The critical theory
is described by two independent level $k=1$ Wess-Zumino-Witten theories
with gauged Lagrangian density
\begin{eqnarray}
{\cal L}_0
\ =&&\
{\rm v_F}\
\bar u\
{\rm i}\gamma^{\mu}
\left(
\partial^{\ }_{\mu}+{\rm i}{\cal A}^e_{\mu}
\right)
\ u
\nonumber\\
+&&\
{\rm v_F}\
\bar v\
{\rm i}\gamma^{\mu}
\left(
\partial^{\ }_{\mu}+{\rm i}{\cal A}^o_{\mu}
\right)
\ v.
\end{eqnarray}
The action has a local color SU(2)$\times$SU(2) chiral symmetry.
There exists within each chain an irrelevant perturbation due to
staggered fluctuations
\begin{eqnarray}
{\cal L}^{\ }_1
\ =&&\
-{\rm v_F}
\left[
\bar u\
{\rm i}\gamma^5
\phi^e_0
\ u
\ +\
\bar u\ u\ \rho^e
\ +\
{{\rm Re}\ E\over2}\ (\rho^e)^2
\right]
\nonumber\\
-&&\
{\rm v_F}
\left[
\bar v\
{\rm i}\gamma^5
\phi^o_0
\ v
\ +\
\bar v\ v\ \rho^o
\ +\
{{\rm Re}\ E\over2}\ (\rho^o)^2
\right].
\end{eqnarray}

Besides the enlarged gauge symmetry,
our critical field theory for the two chains has
a new feature compared to the single chain problem.
There exists an additional global U(2) {\it flavor}
symmetry.
For example, the transformation
$u\rightarrow {1\over\sqrt{2}}(u+v)$,
$v\rightarrow {1\over\sqrt{2}}(u-v)$,
leaves  ${\cal L}_0+{\cal L}_1$ unchanged.
Interactions due to the frustration ${J_1\over J_2}$
break this flavor symmetry
and dynamical off diagonal mass generation
can induce dimerization, or other types of order
\cite{Mudry 1994}. This situation is not unlike the one we encountered
in our study of the frustrated Heisenberg model on a square lattice,
which, in the limit of very small ${J_1\over J_2}$, resembles
two weakly coupled unfrustrated planar antiferromagnet \cite{Mudry 1994}.
Since flavor U(2) has four generators, we expect that there will be a
competition between four independent interactions to destroy criticality.
In particular, dynamical
mass generation in the channel $u^{* }_{a}\ \delta^{\ }_{ab}\ v^{\ }_b$
induces dimerization.
A difference between this field theory and the one close to the limit
${J_2\over J_1}=0$ is that any mean-field gap triggered by the
frustration ${J_1\over J_2}$ cannot be removed anymore by a simple
shift of integration variables  as we did in Eq. (\ref{shiftintegrationvar}).
We leave it to future work for a more detailed study of this theory.

\section{Conclusions}
\label{sec:concl}

The concept of Luttinger liquid, which appears to apply
to a large class of interacting one dimensional electronic systems,
is characterized by the striking phenomenon
of spin and charge separation.
Our goal has been to understand the relationship
between this phenomenon and the separation of spin and charge
quantum numbers predicted
by some mean-field theories for slave holons and spinons.
Here, the slave holon and spinon refer to a picture of the electron in term
of a {\it local} bound state of a pair of holon and spinon
with the electronic spin and charge quantum numbers
carried separately by the spinon and holon, respectively.

Whereas it is easy to show that mean-field theories for slave spinons and
holons do not describe Luttinger liquids, there have been attempts
to recover the properties of the Luttinger liquid
by including {\it perturbatively} fluctuations of the gauge fields.
We have shown explicitly in this paper how,
starting from a mean-field theory for a short-range RVB spin liquid,
one recovers the spin-${1\over2}$ sector of a Luttinger liquid.
A sufficient {\it and} necessary condition for this reconstruction
is to include the {\it strong} fluctuations of the gauge fields constraining
the spinons and to treat them non-perturbatively.

In the case of the frustrated Heisenberg chain for spin-${1\over2}$,
the quantum critical fixed point is described in the language of the
slave spinons by a gauged Wess-Zumino-Witten theory for the group
U(2)/SU(2).
Since this quantum field theory is
equivalent to a level $k=1$ Wess-Zumino-Witten theory
constructed from currents obeying a level $k=1$ SU(2) Kac-Moody
algebra, and having shown that these currents are the infinitesimal
generators of the spin symmetry of the Heisenberg model,
we have recovered Affleck and Haldane's description of
quantum criticality in the spin-${1\over2}$ chain.
This equivalence is a quantum field theory implementation
of the {\it Gutzwiller} projection.
The non-perturbative effects of the gauge fields are to wipe out
any spurious mean-field gap and to eliminate altogether the spinons
from the spectrum by reducing the Virasoro central charge
at mean-field from $C_1=2$ to the physical Virasoro central charge
$C_0=1$.

Umklapp processes for the spin-${1\over2}$
topological excitations of Luttinger liquids
can be relevant perturbations. We have shown that
Umklapp processes for the slave spinons
can also be relevant perturbations to the
quantum critical fixed point.
Our description of the onset of dimerization in
the slave spinon representation makes it clear that one needs to treat
on an equal footing the Affleck-Marston and Anderson order parameters
in order to detect instabilities of the Luttinger liquid.

The essence of the failure of a slave boson (fermion) scheme
to capture some sort of separation of spin and charge
in any ``simple'' way
(say at the Gaussian level around a given mean-field Ansatz)
is due to the fact
that the lower critical space-time dimension for
the deconfinement transition in a pure gauge symmetry
with discrete symmetry is 3. On the other hand, it is not known
if the mechanism for spin and charge separation in Luttinger liquids
generalizes to higher dimensions. If it does, it will be unrelated
to the mechanism of deconfinenent of slave holons and spinons.

\section*{\acknowledgements}

This work was supported in part by NSF grants No.~ DMR91-22385 at the
Department of Physics of the University of Illinois at Urbana-Champaign,
and DMR89-20538 at the Materials Research Laboratory of the University of
Illinois.

\appendix
\section{Uniform and staggered magnetizations}
\label{sec:magnetization}

In this appendix, we want to express the uniform and staggered
magnetizations in terms of the Dirac spinons $u$.
The uniform and staggered magnetization are defined by
\begin{equation}
\vec M^{\ }_{\pm i}
\ =\
{1\over2}(\vec S^{\ }_i\pm\vec S^{\ }_{i+1}),
\quad i\in\Lambda^{\ }_e.
\end{equation}
By combining Eq. (\ref{spininpsibasis}) and the gauge transformation
Eq. (\ref{gaugetrsfkillsutwo}),
one verifies that the uniform magnetization becomes
\begin{eqnarray}
&&
\vec M^{\ }_{+i}\ \propto\
{1\over4}
\pmatrix
{
+
(
u^{\dag}_{\alpha}{\epsilon^{\alpha\beta}\over2}u^{\dag}_{\beta}
+
{\rm H.c.}
)
\cr
-{\rm i}
(
u^{\dag}_{\alpha}{\epsilon^{\alpha\beta}\over2}u^{\dag}_{\beta}
-{\rm H.c.}
)
\cr
+
u^{\dag}_{\alpha}\delta^{\alpha\beta}u^{\ }_{\beta}
},
\end{eqnarray}
while the staggered magnetization becomes
\begin{eqnarray}
&&
\vec M^{\ }_{-i}\propto
{1\over4}
\pmatrix
{
+
(
u^{\dag}_{\alpha}{\rm i}\gamma^1{\epsilon^{\alpha\beta}\over2}u^{\dag}_{\beta}
+
{\rm H.c.}
)
\cr
-{\rm i}
(
u^{\dag}_{\alpha}{\rm i}\gamma^1{\epsilon^{\alpha\beta}\over2}u^{\dag}_{\beta}
-{\rm H.c.}
)
\cr
+
u^{\dag}_{\alpha}{\rm i}\gamma^1\delta^{\alpha\beta}u^{\ }_{\beta}
}.
\end{eqnarray}
Here,
\begin{equation}
\epsilon^{\alpha\beta}\ =\
\pmatrix
{
0&-1\cr
+1&0\cr
}
\ =\
-\epsilon^{\ }_{\alpha\beta}.
\label{epsilontensor}
\end{equation}

The uniform magnetization has a very simple decomposition
with respect to the two chiral sectors:
\begin{equation}
\vec M^{\ }_{+i}\propto
{1\over2}
\left(
\vec {\cal J}^{\ }_+
+
\vec {\cal J}^{\ }_-
\right),
\label{uniformmagnetization}
\end{equation}
where
\begin{equation}
\vec {\cal J}^{\ }_{\pm}\
\ =\
{1\over2}
\pmatrix
{
-(u^{\dag}_{{\pm}1}u^{\dag}_{{\pm}2}+u^{\   }_{{\pm}2}u^{\   }_{{\pm}1})\cr
+{\rm i}(u^{\dag}_{{\pm}1}u^{\dag}_{{\pm}2}
-u^{\   }_{{\pm}2}u^{\   }_{{\pm}1})\cr
+(u^{\dag}_{{\pm}1}u^{\   }_{{\pm}1}+u^{\dag}_{{\pm}2}u^{\   }_{{\pm}2})\cr
}.
\end{equation}
On the other hand, the staggered magnetization mixes the two chiral sectors:
\begin{equation}
\vec M^{\ }_{-i}\propto
{1\over2}
\left(
\vec {\cal K}^{\ }_{+-}
+
\vec {\cal K}^{\ }_{-+}
\right),
\label{staggeredmagnetization}
\end{equation}
where
(
$\vec {\cal K}^{\ }_{-+}$
is obtained from
$\vec {\cal K}^{\ }_{+-}$
by exchanging $-$ and $+$
)
\begin{equation}
\vec {\cal K}^{\ }_{+-}
\ =\
{1\over2}
\pmatrix
{
-
(
u^{\dag}_{+1}u^{\dag}_{-2}
+
u^{\   }_{-2}u^{\   }_{+1}
)
\cr
+{\rm i}
(
u^{\dag}_{+1}u^{\dag}_{-2}
-
u^{\   }_{-2}u^{\   }_{+1}
)
\cr
+
(
u^{\dag}_{+1}u^{\ }_{-1}
+
u^{\dag}_{+2}u^{\ }_{-2}
)
}.
\end{equation}
Here, we have chosen the chiral basis to be
\begin{equation}
\gamma^0=+\tau^2,
\quad
\gamma^1=-{\rm i}\tau^1,
\quad
\gamma^5=-\tau^3.
\end{equation}

\section{Decoupling of the gauge fields from the spinons}
\label{sec:decouplinggauge}

The Lagrangian density of Eq. (\ref{QCDaction}) has two
U(2)$=$U(1)$\times$SU(2) symmetry: the axial and vector
symmetry.
The U(1) symmetry are global:
\begin{eqnarray}
[{\rm U(1)}]_{\ }:\quad
&&
u^{\ } \rightarrow\ ^{\theta}u^{\ }\  =\
e^{+{\rm i}\theta}\ u^{\ },
\nonumber\\
&&
{\cal A}^{\ }_\mu\rightarrow\
^{\theta}{\cal A}^{\ }_\mu\ =\ {\cal A}^{\ }_\mu,
\\
{[{\rm U(1)}]_5}:\quad
&&
u^{\ } \rightarrow\ ^{\theta_5}u^{\ }\  =\
e^{+{\rm i}\theta_5\gamma^{\   }_5}\ u^{\ },
\nonumber\\
&&
{\cal A}^{\ }_\mu\rightarrow\
^{\theta_5}{\cal A}^{\ }_\mu\ =\ {\cal A}^{\ }_\mu.
\end{eqnarray}
The SU(2) symmetry are local:
\begin{eqnarray}
&&
[{\rm SU(2)}]_{\ }:\quad
u^{\ } \rightarrow\
^{\omega}u^{\ }\  =\
U^{\ }_{\omega}\ u^{\ },
\nonumber\\
&&
{\cal A}^{\ }_\mu\rightarrow\
^{\omega}{\cal A}^{\ }_\mu\ =\
U^{\ }_{\omega}\ {\cal A}^{\ }_\mu\ U^{-1}_{\omega}+
({\rm i}\partial^{\ }_\mu\ U^{\ }_{\omega})\ U^{-1}_{\omega},
\\
&&
\protect{[{\rm SU(2)}]_5}:\quad
u^{\ } \rightarrow\
^{\omega_5}u^{\ }\  =\
U^{\ }_{\omega_5}\ u^{\ },
\nonumber\\
&&
{\cal A}^{\ }_\mu\rightarrow\
^{\omega_5}{\cal A}^{\ }_\mu\ =\
U^{\ }_{\omega_5}\ {\cal A}^{\ }_\mu\ U^{-1}_{\omega_5}+
({\rm i}\partial^{\ }_\mu\ U^{\ }_{\omega_5})\ U^{-1}_{\omega_5}.
\end{eqnarray}
The quantum theory does not possess the full gauge invariance.
The continuity equations for the vector and axial currents
\begin{eqnarray}
&&
j^{\mu}\   =\
\bar u^{\ }\                                    \gamma^\mu\  u^{\ },
\quad
\vec J^{\mu}\   =\
\bar u^{\ }\ {\vec\sigma\over2}\                \gamma^\mu\  u^{\ },
\\
&&
j^{\mu}_5\   =\
\bar u^{\ }\                     \gamma^{\   }_5\gamma^\mu\  u^{\ },
\quad
\vec J^{\mu}_5\ =\
\bar u^{\ }\ {\vec\sigma\over2}\ \gamma^{\   }_5\gamma^\mu\  u^{\ },
\end{eqnarray}
cannot be satisfied simultaneously at the quantum level.
This is so because it is impossible to construct a fermionic measure
for the partition function which is simultaneously invariant
under vector and axial gauge transformation \cite{Fujikawa}.
We choose to work with a fermionic measure which is gauge invariant.
We then use the unique property of two space-time dimensions that
allows for the decoupling of the gauge sector from the spinon sector
through a mixture of a vector and axial gauge transformation.

To carry out this program,
it is advantageous to rewrite the Lagrangian density in terms of
the light-cone coordinates:
\begin{equation}
x^{\pm}\ =\ x^0\ \pm\ x^1,\quad
x_{\pm}\ =\ x_0\ \pm\ x_1,
\end{equation}
the Weyl spinors:
\begin{equation}
 u^{\ }_{\pm}\ =\ {1\over2}(1\mp\gamma^{\   }_5)\  u^{\ },
\label{weylspinon}
\end{equation}
and the gauge fields light-cone components
\begin{equation}
{\cal A}^{\pm}\ =\ {\cal A}^0\ \pm\ {\cal A}^1,
\quad
{\cal A}^{\ }_{\pm}\ =\ {\cal A}_0\ \pm\ {\cal A}_1.
\end{equation}
In this basis, the chiral basis, the Lagrangian density is
\begin{equation}
{\cal L}^{\ }_0\ =\
 u^{\dag}_+\ {\rm i} D^{\ }_-\  u^{\ }_+\ +\
 u^{\dag}_-\ {\rm i} D^{\ }_+\  u^{\ }_-,
\end{equation}
where
\begin{equation}
D_{\mp}\ =\
\partial^{\ }_{\mp}\ +\ {\rm i} {\cal A}^{\ }_\mp.
\end{equation}
The global symmetry of the Lagrangian density are now
\begin{eqnarray}
{[{\rm U(1)}]_+}:\quad
&&
 u^{\ }_+ \rightarrow\
^{\theta_+} u^{\ }_+\  =\
e^{+{\rm i}\theta_+}\  u^{\ }_+,
\nonumber\\
&&
A_-\rightarrow\ ^{\theta_+}A_-\ =\ A_-,
\label{chiralu1tsfI}
\\
{[{\rm U(1)}]_-}:\quad
&&
 u^{\ }_- \rightarrow\
^{\theta_-} u^{\ }_-\  =\
e^{+{\rm i}\theta_-}\  u^{\ }_-,
\nonumber\\
&&
A_+\rightarrow\ ^{\theta_-}A_+\ =\ A_+,
\label{chiralu1tsfII}
\end{eqnarray}
whereas the local symmetry of the Lagrangian density are
\begin{eqnarray}
&&
{[{\rm SU(2)}]_+}:\quad
 u^{\ }_+ \rightarrow\
^{\omega_+} u^{\ }_+\  =\
U^{\ }_{\omega_+}\  u^{\ }_+,
\nonumber\\
&&
A^{\ }_-\rightarrow\
^{\omega_+}A^{\ }_-\ =\
U^{\ }_{\omega_+}\ A^{\ }_-\ U^{-1}_{\omega_+}+
({\rm i}\partial^{\ }_-\ U^{\ }_{\omega_+})\ U^{-1}_{\omega_+},
\label{chiralgaugetsfI}
\\
&&
{[{\rm SU(2)}]_-}:\quad
 u^{\ }_- \rightarrow\
^{\omega_-} u^{\ }_-\  =\
U^{\ }_{\omega_-}\  u^{\ }_-,
\nonumber\\
&&
A^{\ }_+\rightarrow\
^{\omega_-}A^{\ }_+\ =\
U^{\ }_{\omega_-}\ A^{\ }_+\ U^{-1}_{\omega_-}+
({\rm i}\partial^{\ }_+\ U^{\ }_{\omega_-})\ U^{-1}_{\omega_-}.
\label{chiralgaugetsfII}
\end{eqnarray}
The classical Noether currents are
\begin{eqnarray}
&&
j_+\ =\
2 u^{\dag}_+\  u^{\ }_+,
\quad
j_-\ =\
2 u^{\dag}_-\  u^{\ }_-,
\label{singletNoether}
\\
&&
\vec J^{\ }_+\ =\
  u^{\dag}_+\ \vec\sigma\  u^{\ }_+,
\quad
\vec J^{\ }_-\ =\
  u^{\dag}_-\ \vec\sigma\  u^{\ }_-.
\label{colorNoether}
\end{eqnarray}

The first step towards decoupling the gauge fields from the spinons
consists in parametrizing the gauge fields
${\cal A}^{\ }_{\mp}$ of the Lie algebra by fields $G^{\ }_{\mp}$
of the Lie group according to
\begin{eqnarray}
&&
{\cal A}^{\ }_-\ =\
+{\rm i}(\partial^{\ }_- G^{\ }_-)G^{-1}_-,
\\
&&
{\cal A}^{\ }_+\ =\
+{\rm i}(\partial^{\ }_+ G^{\ }_+)G^{-1}_+.
\end{eqnarray}
The advantage of such a parametrization is that the transformation law
obeyed by the fields $G^{\ }_{\mp}$ under local
[SU(2)]$_+$ $\times$ [SU(2)]$_-$ transformations
is very simple, namely it amounts to multiplication from the left:
\begin{eqnarray}
&&
G^{\ }_-\ \rightarrow\
^{\omega_+}G^{\ }_-\ =
U_{\omega_+} G_-,
\\
&&
G^{\ }_+\ \rightarrow\
^{\omega_-}G^{\ }_+\ =
U_{\omega_-} G_+.
\end{eqnarray}
The disadvantage is that the relationship between
${\cal A}^{\ }_{\mp}\in$ su(2) and $G^{\ }_{\mp}\in$ SU(2) is non-linear.
It is also important to notice that the
relationship is not one to one since multiplication
from the right of any pair of solutions $G^{\ }_{\mp}$
by a pair of  SU(2) valued matrices which do not
depend on $x^{\mp}$, respectively, is also an appropriate
parametrization. Naturally, one must account for a bosonic Jacobian
when one goes from the su(2) to the SU(2) measures:
\begin{equation}
{\cal D}\ [A_-;A_+]\ =
{\cal D}\ [G_-]\
{\rm Det}\
\left(
\nabla^{\ }_-
\right)\
{\cal D}\ [G_+]\
{\rm Det}\
\left(
{
\nabla^{\ }_+
}
\right),
\end{equation}
where
\begin{equation}
\nabla^{\ }_{\mp}\ \cdot\ =\
\partial^{\ }_{\mp}\ \cdot\
+\
{\rm i}\
[\ +{\rm i}(\partial^{\ }_{\mp} G^{\ }_{\mp})\  G^{-1}_{\mp}\ ,\ \cdot\ ]
\end{equation}
are the covariant derivatives in the adjoint representation.

Clearly, the representation of the gauge fields in terms of the
Lie group valued fields relies on the assumption that the
bosonic Jacobian
${\rm Det}\ (\nabla^{\ }_-)$
${\rm Det}\ (\nabla^{\ }_+)$
is non-vanishing. In other words, we cannot allow for vanishing
eigenvalues of the covariant derivative in the adjoint representation.
This condition restricts the allowed gauge configurations to those
for which the Dirac operator has no zero modes.
The determinants of the covariant derivatives can be converted
into Grassmann integrals
\begin{eqnarray}
&&
{\rm Det}(\nabla_-)\ =\
\int{\cal D}\ [\beta^{\ }_+,\alpha^{\ }_+]\
e^{
+{\rm i}\int\ dx^+dx^-\
{1\over2}{\rm tr}(\beta^a_+\sigma^a\ {\rm i}\nabla_- \sigma^b\alpha^b_+)
},
\\
&&
{\rm Det}(\nabla_+)\ =\
\int{\cal D}\ [\beta^{\ }_-,\alpha^{\ }_-]\
e^{
+{\rm i}\int\ dx^+dx^-\
{1\over2}{\rm tr}(\beta^a_-\sigma^a\ {\rm i}\nabla_+ \sigma^b\alpha^b_-)
}.
\end{eqnarray}
The $\beta$'s and $\alpha$'s are the ghosts needed to fix the gauge.
The ghosts must transform according to
\begin{eqnarray}
{[{\rm SU(2)}]_+}:\quad&&
\alpha^{\ }_+\ \rightarrow\
^{\omega_+}\alpha^{\ }_+\ \ =\
U^{\ }_{\omega_+}\
\alpha^{\ }_+\
U^{-1}_{\omega_+},
\nonumber\\
&&
\beta^{\ }_+\ \rightarrow\
^{\omega_+}\beta^{\ }_+\ =\
U^{\ }_{\omega_+}\
\beta^{\ }_+\
U^{-1}_{\omega_+},
\\
{[{\rm SU(2)}]_-}:\quad&&
\alpha^{\ }_-\ \rightarrow\
^{\omega_-}\alpha^{\ }_-\ \ =\
U^{\ }_{\omega_-}\
\alpha^{\ }_-\
U^{-1}_{\omega_-},
\nonumber\\
&&
\beta^{\ }_-\ \rightarrow\
^{\omega_-}\beta^{\ }_-\ =\
U^{\ }_{\omega_-}\
\beta^{\ }_-\
U^{-1}_{\omega_-},
\end{eqnarray}
since the bosonic measure has the full SU(2)$_+\times$SU(2)$_-$ symmetry
of the Lagrangian density.

The second step is to perform the local gauge transformation
\begin{eqnarray}
&&
 u^{\ }_+\ \rightarrow\
\tilde u^{\ }_+\ =
G^{-1}_+ u^{\ }_+,
\nonumber\\
&&
 u^{\ }_-\ \rightarrow\
\tilde u^{\ }_-\ =
G^{-1}_+ u^{\ }_-,
\nonumber\\
&&
G^{\ }_-\ \rightarrow\
\tilde G^{\ }_-\ =\
G^{-1}_+G^{\ }_-,
\nonumber\\
&&
G^{\ }_+\ \rightarrow\
\tilde G^{\ }_+\ =\
G^{-1}_+G^{\ }_+\ =\ 1,
\end{eqnarray}
under which the $\psi^{\ }_+$'s, $\beta^{\ }_+$'s and $\alpha^{\ }_+$'s
decouple from the bosonic sector. Since the fermionic measure is invariant
under this gauge transformation one can integrate over the fields
$G^{\ }_+$. We choose the factor ${\cal V}^{-1}$ of
Eq. (\ref{QCDpartitionfunction}) to be the gauge volume,
and we are left with the partition function
\begin{equation}
Z\ =\
\int\
{\cal D}\ [\tilde G^{\ }_-]\
\int\
{\cal D}\
[\tilde\beta^{\ }_+,\tilde\alpha^{\ }_+;
 \tilde\beta^{\ }_-,\tilde\alpha^{\ }_-]\
\int\
{\cal D}\
[\tilde u^{\dag}_+,\tilde u^{\ }_+;
 \tilde u^{\dag}_-,\tilde u^{\ }_-]\
e^{+{\rm i}\ S}.
\end{equation}

The third step is to apply the mixed vector and axial gauge
transformation
\begin{eqnarray}
&&
\tilde u^{\ }_+\ \rightarrow\
 u'_+\ =\
\tilde G^{-1}_-\tilde u^{\ }_+\ =\
G^{-1}_- u^{\ }_+,
\nonumber\\
&&
\tilde u^{\ }_-\ \rightarrow\
 u'_-\ =\
\tilde u^{\ }_-\ =\
G^{-1}_+ u^{\ }_-,
\nonumber\\
&&
\tilde G^{\ }_-\ \rightarrow\
G'^{\ }_-\ =\
\tilde G^{-1}_-\tilde G^{\ }_-\ =\ 1,
\nonumber\\
&&
\tilde G^{\ }_+\ \rightarrow\
G'^{\ }_+\ =\
\tilde G^{\ }_+\ =\ 1.
\end{eqnarray}
As promised, it fully decouples the fermions (spinons {\it and} ghosts)
from the bosonic degrees of freedom. However, due to the axial component
of the transformation, the spinon measure {\it and} the ghost measure
change by a non-trivial Jacobian \cite{Polyakov,Karabali 1990,Tanii 1990}:
\begin{eqnarray}
&&
Z\ =\
\int
{\cal D}\mu_b
\int
{\cal D}\mu_f
\
e^{+{\rm i}(S_1+S_2+S_3)},
\label{bosonizedqcdpartition}
\\
&&
{\cal D}\mu_b\ =\
{\cal D}[\tilde G^{\ }_-],
\label{bosonicmeasure}
\\
&&
{\cal D}\mu_f\ =\
{\cal D}
[
\beta'^{\ }_+,\alpha'^{\ }_+;
\beta'^{\ }_-,\alpha'^{\ }_-
]\
{\cal D}\
[
 u'^{\dag}_+, u'_+;
 u'^{\dag}_-, u_-
].
\label{fermionicmeasure}
\end{eqnarray}
The action is the sum of three independent sectors.
The first sector is the sector for free Dirac spinons
\begin{equation}
S_1\ =\
\int\ {dx^+dx^-\over2}\
\big(
 u'^{\dag}_+\ {\rm i}\partial^{\ }_-\  u'_+
\ +\
 u'^{\dag}_-\ {\rm i}\partial^{\ }_+\  u'_-
\big),
\label{freediracspinonaction}
\end{equation}
where the relationship between the free spinons and the original spinons is
\begin{eqnarray}
&&
u'_+\ =\
\tilde G^{-1}_-\ \tilde u^{\ }_+\ =\
G^{-1}_-\  u^{\ }_+,
\label{freerightspinon}
\\
&&
u'_-\ =\
\tilde u^{\ }_-\ =\
G^{-1}_+\  u^{\ }_-.
\label{freeleftspinon}
\end{eqnarray}
The second sector results from the non-invariance of the fermionic
measure under an axial transformation and is given by the
Wess-Zumino-Witten action \cite{Witten 1984}
with {\it negative} level $-1-2c_v$
\cite{onthelevel}
\begin{equation}
S_2\ =\
-(1+2c_v)\ W^{\ }_-[\tilde G^{\ }_-].
\label{negWesszuminowittenaction}
\end{equation}
Here, the Wess-Zumino-Witten action $W^{\ }_-$ depends on the gauge fields
through the non-linear relation
\begin{equation}
\tilde G^{\ }_-\ =\ G^{-1}_+\ G^{\ }_-,
\label{wesszuminowittenfield}
\end{equation}
and is given by
\begin{eqnarray}
W^{\ }_-[G]\ =&&\
+{1\over8\pi}\
\int\ dx^0dx^1\
{\rm tr}
\left[
\partial^{\ }_{\mu}\ G^{\ }\
\partial^{\mu}     \ G^{-1}
\right]
\nonumber\\
\ +&&
{1\over12\pi}\
\int_{ {\rm B},\ {\rm B}=\partial {\rm S}^2 }\
dx^0dx^1dx^2\
\epsilon^{\mu\nu\lambda}\
{\rm tr}
\left[
(\partial^{\ }_{\mu}\ G)G^{-1}\
(\partial^{\ }_{\nu}\ G)G^{-1}\
(\partial^{\ }_{\lambda}\ G)G^{-1}\
\right].
\label{Wesszuminowittenaction}
\end{eqnarray}
Finally, the third sector is the ghost sector
\begin{equation}
S_3\ =\
\int\ {dx^+dx^-\over2}\
\Big[
{\rm tr}\ (\beta'^a_+\sigma^a\ {\rm i}\partial^{\ }_- \sigma^b\alpha'^b_+)
+
{\rm tr}\ (\beta'^a_-\sigma^a\ {\rm i}\partial^{\ }_+ \sigma^b\alpha'^b_-)
\Big].
\label{ghostaction}
\end{equation}

The partition function Eq. (\ref{bosonizedqcdpartition})
describes a two dimensional conformal field theory \cite{Belavin 1984}
with the conformal weights \cite{Belavin 1984,Knizhnik 1984}
\begin{eqnarray}
&&
\left(
\Delta^{\ }_+
,
\Delta^{\ }_-
\right)\left({u'_+}\right)
\ =\
\left(
{1\over2}
,
0
\right),
\nonumber\\
&&
\left(
\Delta^{\ }_+
,
\Delta^{\ }_-
\right)\left({u'_-}\right)
\ =\
\left(
0
,
{1\over2}
\right),
\label{primaryconformalweight}
\\
&&
\left(
\Delta^{\ }_+
,
\Delta^{\ }_-
\right)\left({\tilde G^{\ }_-}\right)
\ =\
\left[
{
{N^2-1\over2N}
\over
N+k
}
\right]_{{N=2\atop k=-1-2N}}\
\left(
1
,
1
\right)
\ =\
\left(
-{1\over4}
,
-{1\over4}
\right),
\nonumber
\end{eqnarray}
for the primary fields $u'$ and $\tilde G^{\ }_-$.

\section{Sugawara construction for a level one spin SU(2) Kac-Moody algebra}
\label{sec:sugawara}

We consider the action Eq. (\ref{fermionsector})
where the $'$ over the Dirac spinons
has been dropped for brevity.
We choose canonical quantization in a box of length $L$ and we impose
periodic boundary conditions.
All products of operators are at equal time.
The spatial coordinate is denoted $x$.
The Fourier convention is
\begin{equation}
u^{\ }_{\sigma a x}\ =\
{1\over\sqrt{L}}\sum_{p\in{\cal Z}} e^{-{\rm i}{2\pi\over L}px}
u^{\ }_{\sigma a p},
\end{equation}
where
\begin{equation}
\sigma\ =\ -,+,
\quad
a\ =\ 1,2,
\quad
0\leq x\leq L.
\end{equation}
The only non-vanishing anticommutators are
\begin{eqnarray}
&&
\{
u^{\   }_{\sigma a x }\ ,\ u^{\dag}_{\sigma'a'x'}
\}
\ =\ \delta^{\ }_{\sigma ,\sigma'}\delta^{\ }_{a,a'}\delta^{\ }_{x,x'},
\\
&&
\{
u^{\   }_{\sigma a p }\ ,\ u^{\dag}_{\sigma'a'p'}
\}
\ =\ \delta^{\ }_{\sigma ,\sigma'}\delta^{\ }_{a,a'}\delta^{\ }_{p,p'}.
\end{eqnarray}
The Hamiltonian is
\begin{eqnarray}
H\ &&=\
\int_0^L dx
\left(
u^{\dag}_{+ax}\ {\rm i}\partial^{\ }_1\ u^{\   }_{+ax}
\ -\
u^{\dag}_{-ax}\ {\rm i}\partial^{\ }_1\ u^{\   }_{-ax}
\right)
\\
&&=\
{2\pi\over L}\
\sum_{p\in{\cal Z}}
\ p\
\left(
u^{\dag}_{+ap}\ u^{\   }_{+ap}
\ -\
u^{\dag}_{-ap}\ u^{\   }_{-ap}
\right).
\end{eqnarray}
The momentum operator is
\begin{eqnarray}
P\ &&=\
\int_0^L dx
\left(
u^{\dag}_{+ax}\ {\rm i}\partial^{\ }_1\ u^{\   }_{+ax}
\ +\
u^{\dag}_{-ax}\ {\rm i}\partial^{\ }_1\ u^{\   }_{-ax}
\right)
\\
&&=\
{2\pi\over L}\
\sum_{p\in{\cal Z}}
\ p\
\left(
u^{\dag}_{+ap}\ u^{\   }_{+ap}
\ +\
u^{\dag}_{-ap}\ u^{\   }_{-ap}
\right).
\end{eqnarray}
The light-cone components of the Hamiltonian and momentum operators are
\begin{eqnarray}
\Theta^{\ }_+\ &&=+
\int_0^L dx\
u^{\dag}_{+ax}\ {\rm i}\partial^{\ }_1\ u^{\   }_{+ax}
\ =+
{2\pi\over L}\
\sum_{p\in{\cal Z}}\
\ p\
u^{\dag}_{+ap}\ u^{\   }_{+ap},
\\
\Theta^{\ }_-\ &&=-
\int_0^L dx\
u^{\dag}_{-ax}\ {\rm i}\partial^{\ }_1\ u^{\   }_{-ax}
\ =-
{2\pi\over L}\
\sum_{p\in{\cal Z}}\
\ p\
u^{\dag}_{-ap}\ u^{\   }_{-ap}.
\end{eqnarray}
The ground state of the Hamiltonian is the Fermi sea
\begin{equation}
|\Psi^{\ }_{fs}\rangle\ =\
\left(
\prod_{p\leq0}
u^{\dag}_{+1p}
u^{\dag}_{+2p}
\right)
\left(
\prod_{p\geq0}
u^{\dag}_{-1p}
u^{\dag}_{-2p}
\right)
|0\rangle.
\end{equation}
The only non-vanishing ground-state expectation value
for bilinears in $u$ is
\begin{equation}
\langle
\Psi^{\ }_{fs}|
\
u^{\dag}_{\sigma a (x -\epsilon)}
\
u^{\   }_{\sigma'a'(x'+\epsilon)}
\
|\Psi^{\ }_{fs}
\rangle
\ =\
\delta^{\ }_{\sigma,\sigma'}
\delta^{\ }_{a,a'}\delta^{\ }_{x,x'}
\cases
{
{+{\rm i}\over4\pi\epsilon+{\rm i}0^+}&if $\sigma=+$,\cr
\hfil&\hfil\cr
{-{\rm i}\over4\pi\epsilon-{\rm i}0^+}&if $\sigma=-$,\cr
}
\end{equation}
and its complex conjugate.

Normal ordering with respect to the Fermi sea is denoted
by $:\cdot:$. The commutator of the current
\begin{equation}
u^{\dag}_{+1x}
u^{\   }_{+1x}
\end{equation}
is defined through the point splitting procedure
\begin{equation}
\left[
:
u^{\dag}_{+1x}
u^{\   }_{+1x}
:
\ ,\
:
u^{\dag}_{+1x'}
u^{\   }_{+1x'}
:
\right]
\ =\
\lim_{{\epsilon\rightarrow0\atop\epsilon'\rightarrow0}}
\left[
:
u^{\dag}_{+1(x-\epsilon )}
u^{\   }_{+1(x+\epsilon )}
:
\ ,\
:
u^{\dag}_{+1(x'-\epsilon')}
u^{\   }_{+1(x'+\epsilon')}
:
\right].
\end{equation}
Wick theorem allows to express the right-hand side solely
in terms of product of (possibly singular) complex functions and
normal ordered products:
\begin{equation}
\left[\
:
u^{\dag}_{+1x}
u^{\   }_{+1x}
:
\ ,\
:
u^{\dag}_{+1x'}
u^{\   }_{+1x'}
:
\right]
\ =\
+{{\rm i}\over\pi}\ \delta'_{x,x'}.
\end{equation}
Here,
the spatial derivative (with respect to $x-x'$) of the delta function
is
\begin{equation}
\delta'_{x,x'}\ =\
\lim_{{\epsilon\rightarrow0\atop\epsilon'\rightarrow0}}
{1\over2(\epsilon +\epsilon')}
\left(
\delta^{\ }_{x-x'+\epsilon+\epsilon',0}
-
\delta^{\ }_{x-x'-\epsilon-\epsilon',0}
\right).
\end{equation}
The coefficient ${1\over\pi}$ depends on the choice of our
conventions. Other commutators of interest
(see appendix
\ref{sec:magnetization}
and
\ref{sec:decouplinggauge}
for the normalization factors of the currents)
are
\begin{eqnarray}
&&
\left[\
:
{j^{\ }_{\sigma  x }\over2}
:
\ ,\
:
{j^{\ }_{\sigma' x'}\over2}
:\
\right]
\ =\
\sigma\ {2{\rm i}\over\pi}
\  \delta^{\ }_{\sigma,\sigma'}\ \delta'_{x,x'},
\\
&&
\left[
\
:
{J^a_{\sigma  x }\over2}
:
\ ,\
:
{J^b_{\sigma' x'}\over2}
:
\
\right]
\ =\
{\rm i}
\epsilon^{abc}
:
{J^c_{\sigma x }\over2}
:
\
\delta^{\ }_{\sigma ,\sigma'}
\
\delta^{\ }_{x ,x'}
\ +\
\sigma {{\rm i}\over2\pi}
\
\delta^{\ }_{\sigma ,\sigma'}
\
\delta'_{x  ,x'},
\label{spinkacmoodyone}
\\
&&
\left[
\
:
{\cal J}^a_{\sigma  x }
:
\ ,\
:
{\cal J}^b_{\sigma' x'}
:
\
\right]
\ =\
{\rm i}
\epsilon^{abc}
:
{\cal J}^c_{\sigma x }
:
\
\delta^{\ }_{\sigma ,\sigma'}
\
\delta^{\ }_{x ,x'}
\ +\
\sigma{{\rm i}\over2\pi}
\
\delta^{\ }_{\sigma ,\sigma'}
\
\delta'_{x  ,x'}.
\label{colorkacmoodyone}
\end{eqnarray}
The factor ${{\rm i}\over\pi}$ comes multiplied by ${\rm tr}\ (\sigma^0)$
and ${\rm tr}\ ({\sigma^0\over4})$, respectively. The algebras of Eqs.
(\ref{spinkacmoodyone}) and (\ref{colorkacmoodyone}) are identical
and equivalent to a level $k=1$ SU(2) Kac-Moody algebra.

Similarly, one can show that
\begin{eqnarray}
\lim_{\epsilon\rightarrow0}\
:
\vec {\cal J}^{\ }_{+(x-\epsilon)}
:
\cdot
:
\vec {\cal J}^{\ }_{+(x+\epsilon)}
:
\ =&&\
+{3\over2}
:
u^{\dag}_{+1x}
u^{\dag}_{+2x}
u^{\   }_{+2x}
u^{\   }_{+1x}
:
+
{3\over4\pi}
:
u^{\dag}_{+x}\
{\rm i}\partial^{\ }_x\
u^{\   }_{+x}
:,
\\
\lim_{\epsilon\rightarrow0}\
:
{\vec J^{\ }_{+(x-\epsilon)}\over2}
:
\cdot
:
{\vec J^{\ }_{+(x+\epsilon)}\over2}
:
\ =&&\
-{3\over2}
:
u^{\dag}_{+1x}
u^{\dag}_{+2x}
u^{\   }_{+2x}
u^{\   }_{+1x}
:
+
{3\over4\pi}
:
u^{\dag}_{+x}\
{\rm i}\partial^{\ }_x\
u^{\   }_{+x}
:,
\\
\lim_{\epsilon\rightarrow0}\
:
{j^{\ }_{+(x-\epsilon)}\over2}
:
:
{j^{\ }_{+(x+\epsilon)}\over2}
:
\ =&&\
+2
:
u^{\dag}_{+1x}
u^{\dag}_{+2x}
u^{\   }_{+2x}
u^{\   }_{+1x}
:
+
{1\over\pi}
:
u^{\dag}_{+x}\
{\rm i}\partial^{\ }_x\
u^{\   }_{+x}
:.
\end{eqnarray}
We have dropped vacuum expectation values on the right-hand side.
Hence, the local generators $T^{\ }_+$
of the energy momentum tensor $\Theta^{\ }_+$ can be expressed
solely in terms of
point split and normal ordered current bilinears:
\begin{eqnarray}
:T^{\ }_+:
\ =&&\
{\pi\over2}
:{j^{\ }_+\over2}:\
:{j^{\ }_+\over2}:\
+\
{2\pi\over3}
:{\vec J^{\ }_+\over2}:\
\cdot\
:{\vec J^{\ }_+\over2}:\
\\
=&&\
{2\pi\over3}
:\vec {\cal J}^{\ }_+:\
\cdot\
:\vec {\cal J}^{\ }_+:\
+\
{2\pi\over3}
:{\vec J^{\ }_+\over2}:\
\cdot\
:{\vec J^{\ }_+\over2}:\
,
\end{eqnarray}
where the vacuum expectation values have been dropped on the right-hand side.
The same relation holds in the $-$ chiral sector.
This completes the
Sugawara construction of the energy-momentum tensor in terms of currents.

Notice that one can rewrite
\begin{equation}
{2\pi\over3}
\ =\
{2\pi\over1+2}
\ =\
\left[
{2\pi\over k+c_v}
\right]_{{k=1\atop c_v=2}},
\end{equation}
if one wants to stress the fact that the currents $\vec{\cal J}^{\ }_{\pm}$
and $\vec J^{\ }_{\pm}$ obey a level $k=1$ Kac-Moody SU(2) algebra.

\section{Useful identities}
\label{sec:identities}

We work with the chiral basis
\begin{equation}
\gamma^0\ =\ +       \tau^2
,\quad
\gamma^1\ =\ -{\rm i}\tau^1
,\quad
\gamma^{\ }_5\ =\ -       \tau^3
{}.
\end{equation}
The components of the spinor ${u}$ are ${u}^{\ }_{\sigma a}$
where $\sigma=+,-$ refers to the Lorentz degrees of freedom
and $a=1,2$ refers to the color degrees of freedom.
One has
\begin{equation}
\gamma^{\ }_5\ {u}^{\ }_{\sigma a}\ =\ -\sigma\ {u}^{\ }_{\sigma a}
,\quad a=1,2, \quad \sigma=+,-.
\end{equation}
As usual $\bar{u}$ will denote
${u}^{\dag}\gamma^0$.
We use repeatedly the identity
\begin{equation}
\vec\sigma^{\ }_{ab}
\ \cdot\
\vec\sigma^{\ }_{cd}
\ =\
2
\delta^{\ }_{ad}
\delta^{\ }_{bc}
\ -\
\delta^{\ }_{ab}
\delta^{\ }_{cd}
\end{equation}
when contracting color indices to express quartic interactions in terms
of quadratic forms in
\begin{eqnarray}
K^{\ }_{+-}
\ &&=\
{u}^{* }_{+a}
\
\delta^{\ }_{ab}
\
{u}^{\   }_{-b}
,
\\
j^{\ }_+
\ &&=\
2\
{u}^{* }_{+a}
\
\delta^{\ }_{ab}
\
{u}^{\   }_{+b}
,
\\
\vec J^{\ }_+
\ &&=\
{u}^{* }_{+a}
\
\vec\sigma^{\ }_{ab}
\
{u}^{\   }_{+b}
,
\end{eqnarray}
together with $K^{\ }_{-+}$, $j^{\ }_-$, and $\vec J^{\ }_-$ obtained by
interchanging $+$ for $-$ and $-$ for $+$. The $K$'s and $j$'s are singlets
under vector gauge transformations. The $\vec J$'s transform like the adjoint
of color SU(2) under vector gauge transformations. Only the $j$'s are singlet
under all chiral transformations. Only the $K$'s induce Umklapp processes.
One finds
\begin{eqnarray}
\left(
\bar{u}\
{u}
\right)^2
\ &&=\
-
K^{\ }_{+-}
\
K^{\ }_{+-}
\ -\
{1\over8}
\
j^{\ }_+
\
j^{\ }_-
\ -\
{1\over2}
\
\vec J^{\ }_+
\cdot
\vec J^{\ }_-
\ +\
\left(
+\leftrightarrow-
\right)
,
\label{masssquareI}
\\
\left(
\bar{u}
\
{\rm i}\gamma^{\ }_5\
\vec\sigma
\
{u}
\right)^2
\ &&=\
-
3\
K^{\ }_{+-}
\
K^{\ }_{+-}
\ -\
{3\over8}
\
j^{\ }_+
\
j^{\ }_-
\ +\
{1\over2}
\
\vec J^{\ }_+
\cdot
\vec J^{\ }_-
\ +\
\left(
+\leftrightarrow-
\right),
\label{chiralcolormassI}
\\
\left(
\bar{u}\
{\rm i}\gamma^5\ {u}
\right)^2
\ &&=\
+
K^{\ }_{+-}
\
K^{\ }_{+-}
\ -\
{1\over8}
\
j^{\ }_+
\
j^{\ }_-
\ -\
{1\over2}
\
\vec J^{\ }_+
\cdot
\vec J^{\ }_-
\ +\
\left(
+\leftrightarrow-
\right)
,
\\
\left(
\bar{u}
\
\vec\sigma
\
{u}
\right)^2
\ &&=\
+
3\
K^{\ }_{+-}
\
K^{\ }_{+-}
\ -\
{3\over8}
\
j^{\ }_+
\
j^{\ }_-
\ +\
{1\over2}
\
\vec J^{\ }_+
\cdot
\vec J^{\ }_-
\ +\
\left(
+\leftrightarrow-
\right).
\end{eqnarray}
This results in the important identities
\begin{eqnarray}
&&
\left(
\bar{u}
\
{u}
\right)^2
\ =\
{1\over3}
\left(
\bar{u}
\
{\rm i}\gamma^{\ }_5\
\vec\sigma
\
{u}
\right)^2
\ -\
{2\over3}
\left[
\vec J^{\ }_+\cdot\vec J^{\ }_-
\ +\
(
+\leftrightarrow-
)
\right]
,
\label{masssquareII}
\\&&
\left(
\bar{u}
\
{\rm i}\gamma^5\
{u}
\right)^2
\ =\
{1\over3}
\left(
\bar{u}
\
\vec\sigma
\
{u}
\right)^2
\ -\
{2\over3}
\left[
\vec J^{\ }_+\cdot\vec J^{\ }_-
\ +\
(
+\leftrightarrow-
)
\right]
{}.
\end{eqnarray}
The singlet current-current interaction can be expressed as
\begin{equation}
{1\over8}\
j^{\ }_+
j^{\ }_-
\ =\
2\
\vec {\cal J}^{ }_+\cdot\vec {\cal J}^{ }_-
\ -\
\left(
u^{* }_{+1}u^{\ }_{-1}
\
u^{* }_{+2}u^{\ }_{-2}
\ +\ {\rm H.c.}
\right),
\end{equation}
so that
\begin{equation}
K^{\ }_{+-}
\
K^{\ }_{+-}
\ +\
{1\over8}
\
j^{\ }_+
\
j^{\ }_-
\ =\
2\
\vec {\cal J}^{ }_+\cdot\vec {\cal J}^{ }_-
\ +\
\sum_{a=1,2}
u^{* }_{+ax}u^{\ }_{-ax}
\
u^{* }_{+a(x+\bar\epsilon)}u^{\ }_{-a(x+\bar\epsilon)}
{}.
\end{equation}
With the help of
Eq. (\ref{chiralcolormassI})
we obtain the second important identity
relating the Umklapp interaction with the interactions
$\vec{\cal J}^{\ }_+\cdot\vec{\cal J}^{\ }_-$:
\begin{equation}
\sum_{a=1,2}
\left[
u^{* }_{+ax}
u^{* }_{+a(x+\bar\epsilon)}\
u^{\ }_{-ax}
u^{\ }_{-a(x+\bar\epsilon)}
\ +\
(+\leftrightarrow-)
\right]
\ =\
4\ \vec {\cal J}^{ }_+\cdot\vec {\cal J}^{ }_-
\ +\
{1\over3}
\left(
\bar{u}
\
{\rm i}\gamma^5\
{u}
\right)^2
\ -\
{1\over3}
\vec J^{\ }_+\cdot\vec J^{\ }_-
{}.
\label{umklappidentity}
\end{equation}


\begin{references}

\bibitem{tjmodel}
J. E. Hirsch,
Phys. Rev. Lett. {\bf 54},1317 (1985);
%
P. W. Anderson, Science {\bf 235}, 1196 (1987);
%
C. Gros, R. Joynt and T. M. Rice,
Phys Rev. B {\bf 36}, 381 (1987);
%
F. C. Zhang and T. M. Rice,
Phys. Rev. B {\bf 37}, 3759 (1988).

\bibitem{tJisluttingernum}
S. Sorella, A. Parola, M. Parrinello, and E. Tossatti,
Europhys. Lett. {\bf 12}, 721 (1990);
A. Parola and S. Sorella, Phys. Rev. Lett. {\bf 64}, 1831 (1990);
M. Ogata and H. Shiba, Phys. Rev. B {\bf 41}, 2326 (1990);
C.S. Hellberg and E.J. Mele, Phys. Rev. B {\bf 48}, 646 (1993).

\bibitem{hubbisluttingeranal}
H.J. Schulz, Phys. Rev. Lett. {\bf 64}, 2831 (1990);
N. Kawakami and S.K. Yang, Phys. Lett {\bf 148A}, 359 (1990);
H. Frahm and V.E. Korepin, Phys. Rev. B {\bf 42}, 10553 (1990).

\bibitem{tJisluttingeranal}
N. Kawakami and S.K. Yang, Phys. Rev. Lett. {\bf 65}, 2309 (1990)
and J. of Phys. Cond. Matt. {\bf 3}, 5983 (1991).

\bibitem{Haldane 1980}
F. D. M. Haldane, Phys. Rev. Lett. {\bf 45}, 1358 (1980).

\bibitem{slavefermibose}
S. E. Barnes,
J. Phys. F {\bf 6}, 1375 (1976), and {\bf 7}, 2637 (1977);
%
P. Coleman,
Phys. Rev. B {\bf 29}, 3035 (1984);
%
G. Kotliar and A. E. Ruckenstein,
Phys. Rev. Lett. {\bf 57}, 1362 (1986).

\bibitem{Baskaran 1987}
G. Baskaran, Z. Zou and P. W. Anderson,
Solid State Commun. {\bf 63}, 973 (1987).

\bibitem{Zou 1988}
Z. Zou and P. W. Anderson,
Phys. Rev. B {\bf 37}, 627 (1988).

\bibitem{Arovas 1988}
D. P. Arovas and A. Auerbach,
Phys. Rev. B {\bf 38}, 316 (1988);

\bibitem{Read 1989}
N. Read and B. Chakraborty, Phys. Rev. B {\bf 40}, 7133 (1989).

\bibitem{Flensberg 1989}
K. Flensberg, P. Hedegard, and M.B. Pedersen, Phys. Rev. B {\bf 40}, 850
(1989).

\bibitem{Feng 1993}
S. Feng, J.B. Wu, Z.B. Su, and L. Yu, Phys. Rev. B {\bf 47}, 15192 (1993).

\bibitem{Wen 1991}
X. G. Wen,
Phys. Rev. B {\bf 44}, 2664 (1991).

\bibitem{Mudry 1994}
C. Mudry and E. Fradkin,
Phys. Rev. B {\bf 44}, 5200 (1994).

\bibitem{Wegner 1971}
F. J. Wegner, J. Math. Phys. {\bf 12}, 2259 (1971).

\bibitem{Faddeev 1981}
L. D. Faddeev and L. A. Takhtajan,
Phys. Lett. A {\bf 85}, 375 (1981).

\bibitem{j2/j1=0.5}
C.K. Majumbar, J. Phys. {\bf C3}, 911 (1969);
C.K. Majumbar and D.K. Ghosh, J. Math. Phys. {\bf 10}, 1388, 1399 (1969);
P.M. van den Broek, Phys. Lett. {\bf 77A}, 261 (1980);
B.S. Shastry and B. Sutherland, Phys. Rev. Lett. {\bf 47}, 964 (1981).

\bibitem{Haldane 1982}
F.D.M. Haldane,
Phys. Rev. B {\bf 25}, 4925 (1982).

\bibitem{Ogata 1991}
M. Ogata, M.U. Luchini and T.M. Rice, Phys. Rev. B {\bf 44}, 12083 (1991).

\bibitem{doublechain}
S.P. Strong and A.J. Millis, Phys. Rev. Lett. {\bf 69}, 2419 (1992);
T. Barnes, E. Dagotto and J. Riera, E.S. Swanson, Phys. Rev. B {\bf 47},
3196 (1993);
S. Gopalan, T.M. Rice, and M. Sigrist, cond-mat/9312026;
S.R. White, R.M. Noack, and D.J. Scalapino, cond-mat/9403042.

\bibitem{t-Jladder}
E. Dagotto, J. Riera, and D. Scalapino, Phys. Rev. B {\bf 45}, 5744 (1992);
T.M. Rice, S.Gopalan, and M. Sigrist, Europhys. Lett. {\bf 23}, 445 (1993);
D.V. Khveshchenko and T.M. Rice, cond-mat/9401010;
M. Sigrist, T.M. Rice, and F.C. Zhang, cond-mat/9401019.

\bibitem{Affleck 1987}
I. Affleck and F.D.M. Haldane, Phys. Rev. B {\bf 36}, 5291 (1987).

\bibitem{Affleck 1986}
I. Affleck, Nucl. Phys. {\bf B265}, 409 (1986).

\bibitem{Affleck 1988}
I. K. Affleck and J. B. Marston,
Phys. Rev. B {\bf 37}, 3774 (1988).

\bibitem{Hsu 1988}
I. K. Affleck, Z. Zou, T. Hsu and P. W. Anderson,
Phys. Rev. B {\bf 38}, 745 (1988).

\bibitem{Dagotto 1988}
E. Dagotto, E. Fradkin and A. Moreo,
Phys. Rev. B {\bf 38}, 2926 (1988).

\bibitem{Baskaran 1989}
G. Baskaran, Physica Scripta {\bf T27}, 53 (1989).

\bibitem{particle-hole}
The identities
$
|0>_{s_i}=
\psi^{\dag}_{i2}|0>_{\psi_i}
$,
$
s^{\dag}_{i\uparrow  }|0>_{s_i}=
\psi^{\dag}_{i1}\psi^{\dag}_{i2}|0>_{\psi_i}
$,
$
s^{\dag}_{i\downarrow}|0>_{s_i}=
|0>_{\psi_i}
$,
and
$
s^{\dag}_{i\uparrow  }s^{\dag}_{i\downarrow}|0>_{s_i}=
\psi^{\dag}_{i1}|0>_{\psi_i}
$,
where $|0>_{s_i}$ is the spinon vacuum at site $i$, hold.

\bibitem{pathological caseI}
It could be that in some pathological cases
$a^1_{\hat{\rm o}}$ is such that the gap closes at $\pm{\pi\over2}$
for non-vanishing $E$ and $X$. We never encountered this situation in our
numerics. We will assume that it never happens in practice.

\bibitem{pathological caseII}
We assume that the zeroes of $|\vec\xi^{\ }_k|$ are simple.

\bibitem{Thirring 1958}
W. Thirring, Annals of Phys. {\bf 3}, 91 (1958).

\bibitem{Fradkin 1991}
For a review, see E. Fradkin,
{\it Field Theories of Condensed Matter Systems}
(Addison-Wesley, Redwood City, CA, 1991), chapter 4.

\bibitem{Karabali 1990}
D. Karabali and H. Schnitzer, Nucl. Phys. {\bf B329}, 649 (1990).

\bibitem{Tanii 1990}
Y. Tanii, Modern Phys. Lett. A {\bf 5}, 927 (1990).

\bibitem{Itoi locarno}
C. Itoi and H. Mukaida, unpublished.

\bibitem{GKO}
P. Goddard, A. Kent and D. Olive,
Phys. lett. {\bf 152B}, 88 (1985).

\bibitem{Witten 1984}
E. Witten, Commun. Math. Phys. {\bf 92},
455 (1984).

\bibitem{Polyakov}
A. M. Polyakov and P. M. Wiegmann,
Phys. Lett. {\bf 131B}, 121 (1983), {\it ibid} {\bf 141B}, 223 (1984).

\bibitem{Polyakovhouches}
A. M. Polyakov in {\it Fields, Strings and Critical Phenomena},
edited by E. Br\' ezin and J. Zinn-Justin
Les Houches, Session XLIX, 1988 (Elsevier Science, 1989).

\bibitem{Knizhnik 1984}
V.G. Knizhnik and A.B. Zamolodchikov,
Nucl. Phys. B {\bf 247}, 83 (1984).

\bibitem{Belavin 1984}
A.A. Belavin, A.M. Polyakov and A.B. Zamolododchikov,
Nucl. Phys. {\bf B241}, 333 (1984).

\bibitem{Sugawara 1968}
H. Sugawara, Phys. Rev. {\bf 170}, 1659 (1968);
C. Sommerfield, {\it ibid} {\bf 176}, 2019 (1968);
S. Coleman, D. Gross and R. Jackiw, {\it ibid} {\bf 180}, 1359 (1969).

\bibitem{Schwinger 1959}
J. Schwinger, Phys. Rev. Lett. {\bf 3}, 296 (1959).

\bibitem{QED}
This is the case, for instance, for $QED_2$ in the strong
coupling regime.

\bibitem{Tonegawa 1987}
T. Tonegawa and I. Harada, J. Phys. Soc. Jpn. {\bf 56}, 2153 (1987).

\bibitem{Fujikawa}
K. Fujikawa, Phys. Rev. Lett. {\bf 42}, 1195 (1979);
{\it ibid} Phys. Rev. D {\bf 21}, 2848 (1980).

\bibitem{onthelevel}
The contribution $-1$ to the level of the Wess-Zumino-Witten
action comes from the measure of the spinons while $-2c_v$ comes from
the measure of the ghosts. Here, $c_v$ is the Casimir eigenvalue in
the adjoint representation of SU(2), i.e., $c_v= 1(1+1)$.

\end{references}
\end{document}